\documentclass[graybox]{svmult}
\usepackage{type1cm}        
\usepackage{makeidx}         
\usepackage{graphicx}        
\usepackage{multicol}        
\usepackage[bottom]{footmisc}
\usepackage{newtxtext}       %
\usepackage{newtxmath}       
\usepackage{array,multirow}
\usepackage{csquotes}
\input{qcircuit}
\usepackage{marginnote}
\usepackage{xcolor} 
\makeindex             

\renewcommand\leq{\leqslant}
\def\Tr{\operatorname{Tr}}\def\n#1{|\!|#1|\!|}\def\>{\rangle}\def\<{\langle}
\def\trnsfrm#1{\mathcal #1}
\def\tA{\trnsfrm A}\def\tB{\trnsfrm B}\def\tC{\trnsfrm C} \def\tE{\trnsfrm E}
 
\def\tT{\trnsfrm T}\def\tU{\trnsfrm U}\def\tV{\trnsfrm V} 
\def\tM{\trnsfrm M}\def\tX{\trnsfrm X} 
\def\tO{\trnsfrm O}\def\tE{\trnsfrm E}\def\tR{\trnsfrm R}
\def\tP{\trnsfrm P } \def\tT{\trnsfrm T} 
\def\rA{{\rm A}}\def\rB{{\rm B}}\def\rC{{\rm C}}\def\rD{{\rm D}} \def\rE{{\rm E}} \def\rF{{\rm F}}
\def\rG{{\rm G}}\def\rH{{\rm H}}\def\rI{{\rm I}}\def\rL{{\rm L}}\def\rM{{\rm M}}
\def\rN{{\rm N}}\def\rO{{\rm O}}
 
\def\sH{{\mathcal{H}}}
\def\St{\rm{St}}\def\Eff{\mathrm{Eff}}\def\Trn{\mathrm{Trn}}
\def\Mrkv{\mathrm{Mrkv}}\def\Prm{\mathrm{Prm}}\def\T{\mathrm{T}}\def\Bnd{\mathrm{Bnd}}
\def\Cmplx{\mathbb{C}}\def\Reals{\mathbb{R}}
\def\vx{{\bf x}}
\def\Set{{\mathsf S}}\def\Conv{\mathsf{Conv}}\def\Cone{\mathsf{Cone}}

\def\Uset{{\mathbb U}} 
\def\Supp{\mathsf{Supp}\,}\def\Proj{\mathsf{Proj}\,}
\def\CP{\mathrm{CP}}\def\P{\mathrm{P}}

\def\Zc#1{\prepareC{#1}}				%
\def\ZS#1{\poloFantasmaCn{#1}\qw}	
\def\ZD#1#2{\multimeasureD{#1}{#2}}	
\def\Zd#1{\measureD{#1}}			%
\def\ZG#1#2{\multigate{#1}{#2}}		
\def\Zg#1{\gate{#1}}				%
\def\Zh#1{\ghost{#1}}	

\def\redb{\textcolor{red}{\bullet}}\def\greenb{\textcolor{green}{\bullet}}

\begin{document}

\title*{Hard Problem and Free Will:\\  an information-theoretical approach}
\titlerunning{Hard Problem and Free Will}
\author{Giacomo Mauro D'Ariano and Federico Faggin}
\authorrunning{Giacomo Mauro D'Ariano and Federico Faggin}
\institute{Giacomo Mauro D'Ariano \at Dipartimento di Fisica, 
University of Pavia, via Bassi 6, 27100 Pavia, \email{dariano@unipv.it}
\and Federico Faggin\at Federico and Elvia Faggin Foundation}
\maketitle
\begin{displayquote}
 \begin{center}
{\em We are such stuff as dreams are made on, and our little life is rounded with a sleep. } 
\begin{flushright}
William Shakespeare
\end{flushright}
\end{center}
\end{displayquote}
\abstract{We explore definite theoretical assertions about consciousness, starting from a non-reductive psycho-informational solution of  David Chalmers's  {\em hard problem}, based on the hypothesis that a fundamental property of "information" is its experience by the supporting "system". The kind of information involved in consciousness needs to be quantum for multiple reasons, including its intrinsic privacy and its power of building up thoughts by entangling qualia states. As a result we reach a quantum-information-based panpsychism, with classical physics supervening on quantum physics, quantum physics supervening on quantum information, and quantum information supervening on consciousness. \smallskip\\
  We then argue that the internally experienced quantum state, since it corresponds to a definite experience--not to a random choice--must be pure, and we call it {\em ontic}. This should be distinguished from the state predictable from the outside (i.e. the state describing the knowledge of the experience from the point of view of an external observer) which we call {\em epistemic} and is generally mixed. Purity of the ontic state requires an evolution that is purity preserving, namely a so-called {\em atomic} quantum operation. The latter is generally probabilistic, and its particular outcome is interpreted as the free will, which is unpredictable even in principle since quantum probability cannot be interpreted as lack of knowledge. We also see how the same purity of state and evolution allow solving the well-known {\em combination problem} of panpsychism. 
\smallskip\\
Quantum state evolution accounts for a {\em short-term buffer of experience} and contains itself quantum-to-classical and classical-to-quantum information transfers. Long term memory, on the other hand, is classical, and needs memorization and recall processes that are quantum-to-classical and classical-to-quantum, respectively. Such processes can take advantage of multiple copies of the experienced state re-prepared with "attention", and therefore allowing for a better quality of classical storing. 
\smallskip\\
Finally, we explore the possibility of experimental tests of our theory in cognitive sciences, including the evaluation of the number of qubits involved, the existence of complementary observables, and violations of local-realism bounds. 
\smallskip\\
In the appendices we succinctly illustrate the {\em operational probabilistic theory} (OPT) framework for possible post-quantum theories of consciousness, assessing the convenient black-box approach of the OPT, along with its methodological robustness in separating objective from theoretical elements, guaranteeing experimental control and falsifiability. We finally synthetically compare the mathematical postulates and theorems of the most relevant instances of OPTs--i.e. classical and quantum theories--for convenience of the reader for better understanding our theory of consciousness.
The mathematical notation is provided in a handy table in the appendices.}
\section{A quantum-informational panpsychism}
In his book {\em The Character of Consciousness}     \cite{Chalmers2010} David Chalmers states what he calls  the {\em hard problem of consciousness}, namely the issue of explaining  our {\em experience}--sensorial, bodily, mental, and emotional, including any stream of thoughts. Chalmers contrasts the {\em hard problem} with the {\em easy problems} which, as it happens in all sciences, can be tackled in terms of  a mechanistic approach that is useless for the problem of experience. Indeed, in all sciences we always seek explanations in terms of  {\em functioning}, a concept that is entirely independent from the notion of  {\em experience}. 
Chalmers writes: 
\begin{displayquote}
{\em Why is the performance of these functions accompanied by experience? ...

Why doesn't all of this information processing go on ``in the dark'' free of any inner feel?...

There is an explanation gap between the function and the experience.}
\end{displayquote}
An effective paradigm for comprehending the conceptual gap between ``experience'' and ``functioning'' is that of {\em zombie}, which is behaviourally indistinguishable from a conscious being, nevertheless has no inner experience. 

\bigskip
There are currently two main lines of response to the hard problem: 1) the {\em Physicalist view}--with consciousness ``emergent from a functioning'', such as some biological property of life     \cite{dennett2017consciousness}; 2) the {\em Panpsychist view}--with consciousness as a fundamental feature of the world that all entities have. What is proposed here is: 
\begin{itemize}
\item[] Panpsychism with consciousness as a fundamental feature of  ``information'', and  physics supervening on information.
\end{itemize}

The idea that physics is a manifestation of pure information processing  has been strongly advocated by John Wheeler      \cite{wheeler89} and Richard Feynman      \cite{feynman1982simulating,hey1998feynman}, along with several other authors, among which David Finkelstein     \cite{finkelstein1996quantum}, who was particularly fond of this idea     \cite{PWP}. Only quite recently, however, the new informational paradigm for physics has been concretely established. This program achieved: 1) the derivation of quantum theory  as an information theory    \cite{QUIT-Arxiv,purification,CUPDCP}, and 2) the derivation of free quantum field theory as emergent from the nontrivial quantum algorithm with denumerable systems with  minimal algorithmic complexity    \cite{PhysRevA.90.062106,bisio2016}.\footnote{The  literature on the informational derivation of free quantum field theory is extensive, and, although not up to date, we suggest the review     \cite{PWP} written by one of the authors in memoriam of David Finkelstein. The algorithmic paradigm has opened for the first time the possibility of avoiding physical primitives in the axioms of the physical theory, allowing a re-foundation of the whole physics over logically solid grounds    \cite{DArianoHilbert}.} In addition to such methodological value, the new information-theoretic derivation of quantum field theory is particularly promising for establishing a theoretical framework for quantum gravity as emergent from the quantum information processing, as also suggested by the role played by information in the holographic principle     \cite{Susskind:1995aa,Bousso:2003aa}. In synthesis: the physical world emerges from an underlying algorithm, and the kind of information that is processed beneath is quantum.

\smallskip
The idea that quantum theory (QT) could be regarded as an information theory is a relatively recent one    \cite{fuchs2002quantum}, and originated within the field of {\em quantum information}     \cite{Nielsen:1997p655}. Meanwhile what we name ``information theory'' has largely evolved, from its origins as a communication theory     \cite{Schannon}, toward a general theory of  ``processing'' of information, which previously had been the sole domain of computer science. 

\smallskip
What do we mean by ``information theory''?
\smallskip
 
Recently, both in physics and in computer science (which in the meantime connected with quantum information), the theoretical framework for all information theories emerged in the physics literature in terms of the notion of {\em Operational Probabilistic Theory} (OPT)    \cite{purification,CUPDCP,Chiribella2016}. an isomorphic framework that emerged within computer science in terms of {\em Category Theory}    \cite{CoeckePV,TullPV,Coecke2012,Coecke2016,abramsky_heunen_2016,Coecke2018}.
Indeed, the mathematical framework of an information theory is precisely that of the OPT, whichever information theory we consider--either classical, quantum, or "post-quantum".  The main structure of  OPT is reviewed in the Appendix.\footnote{
The reader who is not familiar with the notion of OPT can regard the OPT  as the mathematical formalization of the rules for building quantum information circuits. For a general idea about OPT it is recommended to read the appendix. The reader is supposed to be familiar with elementary notions, such as {\em state} and {\em transformation} with finite dimensions.}

Among the information theories, classical theory (CT) plays a special role. In fact, besides being itself an OPT,  CT enters the  operational framework in terms of objective outcomes of the theory, which for causal OPTs (as QT and CT) can be used for conditioning the choice of a following transformation within a set. Clearly this also happens in the special case of  QT. Thus, the occurrence of a given outcome can be regarded as a quantum-to-classical information exchange, whereas conditioning constitutes a classical-to-quantum information exchange. We conclude that we should regard the physical world faithfully ruled by both quantum and classical theories together, with information transforming between the two types. 

This theoretical description of reality should be contrasted with the usual view of reality as being quantum, creating a fallacy of misplaced concreteness. The most pragmatic point of view is to regard QT and CT together as the correct information theory to describe reality, without incurring any logical paradox. We will use this idea in the rest of the chapter. Due to the implicit role 
played by CT in any OPT, when we mention CT we intend to designate the special corresponding OPT.

\smallskip
We will call the present view, with consciousness as fundamental for information and physics supervening on quantum information: {\em Quantum-Information Panpsychism}.

\smallskip
In place of QT, one may consider a {\em post-quantum} OPT--e.g. RQT  (QT on real Hilbert spaces), FQT (Fermionic QT), PRB (an OPT build on Popescu-Rohlich boxes    \cite{Popescu1992}), etc.     \cite{CUPDCP}, yet some of the features of the present consciousness theory can be translated into the other OPT, as long as the notion of ``entanglement'' survives in the considered OPT.

\section{A non-reductive psycho-informational solution: general principles}
The fundamental nature of the solution to the hard problem proposed here has been suggested by David Chalmers as satisfying  the following requirements     \cite{Chalmers2010} 
\begin{itemize}
\itemsep=3pt
\item[]Chalm$_1$: {\em Consciousness as fundamental entity, not explained in terms of anything simpler...}
\item[]Chalm$_2$: {\em ... a non reductive theory of experience will specify basic principles that tell us how basic experience depends on physical features of the world.}
\item[]Chalm$_3$: {\em These psychophysical principles will not interfere with physical laws (closure of physics). Rather they will be a supplement to physical theory.}
\end{itemize}

In an information-theoretic framework, in which physics supervenes on information our principle will be {\em psychoinformational} (see Chalm$_3$). The non-reductive (Chalm$_{2}$) psycho-informational principle that is proposed here is the following:
\begin{itemize}
\item[P1:] \hskip4pt psychoinformational principle: {\em Consciousness is the information-system's experience of its own information state and processing.
}
\end{itemize}

As we will see soon, it is crucial that the kind of information that is directly experienced be quantum.

Principle P1 may seem {\em ad hoc}, but the same happens with the introduction of any fundamental quantity (Chalm$_{1,2}$) in physics, e.g. the notions of {\em inertial mass}, {\em electric charge}, etc. Principle P1 asserts that {\em experience is a fundamental feature of information}, hence also of physics, which supervenes on it. P1 is not reductive (Chalm$_2$), and it does not affect physics (Chalm$_3$), since the kind of information involved in physics is  quantum+classical. On the other hand, P1 supplements physics (Chalm$_3$), since the latter supervenes on information theory.

\medskip
It is now natural to ask: {\em which are the systems?} Information, indeed, is everywhere: light strikes objects and thereafter reaches our eyes, providing us with information on those objects: colour, position, shape,... Information is supported by a succession of systems: the light modes, followed by the retina, then the optical nerves, and finally the several bottom-up and top-down visual processes occurring in the brain. Though the final answer may have to come from neuroscience, molecular biology, and cognitive science experiments, we can use the present OPT approach to inspire crucial experiments. OPT has the power of being a {\em black-box} method that does not need the detailed "physical" specification of the systems, and this is a great advantage! And, indeed, our method tackles the problem in terms of a pure in-principle reasonings, independent of the "hardware" supporting information, exactly as we do in information theory. Such {\em hardware independence} makes the approach particularly suited to address a problem so fundamental as the problem of consciousness. 

\medskip
We now proceed with the second principle:

\begin{itemize}
\item[P2:]\hskip4pt Privacy principle: {\em Experience is not sharable, even in principle.}
\end{itemize}
Principle P2 plays a special role in selecting which information theories  are compatible with a theory of consciousness. Our experience is indeed not sharable: this is a fact. We hypothesize that that non shareability of experience holds {\em in principle}, not just as a technological limitation.\footnote{It may be possible  to know which systems are involved in a particular experience, as considered in Ref.    \cite{Loorits2014}. However, we could never know the experience itself, since it corresponds to non sharable quantum information.}A crucial fact is that {\em information shareability} is equivalent to {\em information readability with no disturbance}.\footnote{In fact, information shareability is equivalent to the possibility of making copies of it--technically {\em cloning information}. The possibility of cloning information, in turn, is equivalent to that of {\em reading information without disturbing it}. Indeed, if one could read information without  disturbing it, he could read the information as many times as needed in order to acquire all its different complementary sides, technically performing a {\em tomography} of the information. And once he knows all the information he can prepare copies of it at will. Viceversa, if one could clone quantum information, he could read clones keeping the original untouched.} 
Recently it has been proved  that the only theory where any information can be extracted without disturbance is classical information theory     \cite{tosini2020quantum}. We conclude that P2 implies that a theory of consciousness needs nonclassical information theory, namely QT, or else a ``post-quantum'' OPT. 

Here we will consider the best known instance of OPT, namely QT. As we will see, such a choice of theory turns out to be very powerful in accounting for all the main features of consciousness. We state this choice of theory as a principle: 

\begin{itemize}
\item[P2':]\hskip4pt \hspace{1mm}Quantumness of experience: {\em The information theory of consciousness is quantum theory.}
\end{itemize}
We remind that  classical theory is always accompanying any OPT, thus there will be exchanges and conversions of classical and quantum information.

\begin{figure}[h]
\begin{center}
\includegraphics[width=9cm]{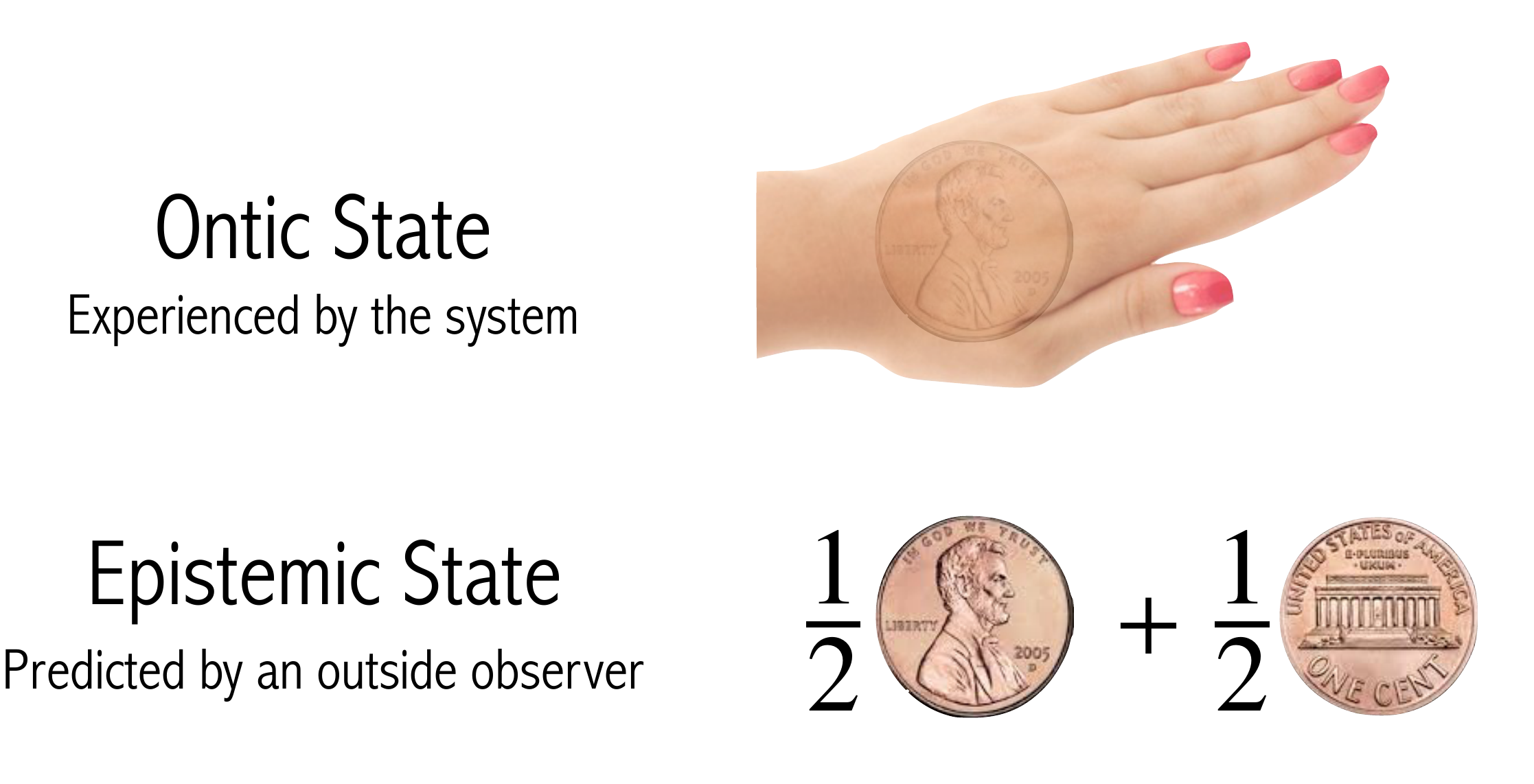}
\caption{\label{fig:flowchart} Illustration of the notions of  {\em ontic} and {\em epistemic} states for a system given by a classical bit, here represented by a coin, with states 0=HEAD or 1=TAIL. We call {\em ontic} the ``actual'' state of the system, which is pure and generally unknown, except as an unsharable experience (in the figure it is the coin state HEAD covered by the hand). We call {\em epistemic} the state that represents the knowledge  about the system of an outside observer, e.g. the state of the unbiased coin corresponds to $\tfrac{1}{2}$HEAD+$\tfrac{1}{2}$TAIL.  
}\label{f:ontic-epistemic}
\end{center}
\end{figure}

We now introduce a third principle:

\begin{itemize}
\item[P3:]\hskip4pt Psycho-purity principle: {\em The state of the conscious system is pure.}
\end{itemize}
P3 may seem arbitrary if one misidentifies the {\em experience} with the {\em knowledge of it}. 
The actual experience is {\em ontic} and {\em definite}.  It is {\em ontic}, \`a la Descartes ({\em Cogito, ergo sum} ...). The existence of our experience is surely something that everybody would agree on, something we can be sure of. It is {\em definite}, in the sense that one has only a single experience at a time--not a probability distribution of experiences. The latter would describe the {\em knowledge of somebody else's experience}. Consider that the experience of a {\em blurred coin} is just as definite as the experience of a {\em sharp coin} since they are two different definite experiences. It is the knowledge of the coin side that is definite, which only occurs if the image of the coin is sufficiently sharp, otherwise it would be represented in probabilistic terms, e.g. by the mixture-state $\tfrac{1}{2}$TAIL+$\tfrac{1}{2}$HEAD, based on a fair-coin hypothesis (see Fig. \ref{f:ontic-epistemic}). We thus can understand that by the same definition the notion of mixture is an  {\em epistemic} one, based on our prior knowledge about the coin having two possible states. In synthesis:
\begin{enumerate}
\item {\em Experience} is described by an {\em ontic quantum state}, which is {\em pure}.
\item {\em Knowledge} is described by an {\em epistemic quantum state}, which is generally {\em mixed}.
\end{enumerate}
We will see later how principle P3 is crucial for solving the {\em combination problem}.
\par Let's now state our fourth principle, which is theory independent:
\begin{itemize}
\item[P4:]\hskip4pt Qualia principle: {\em Experience} is made of  structured qualia. 
\end{itemize}
{\em Qualia} (phenomenal qualitative properties) are the building blocks of  conscious experience. 
Their existence has so far been inexplicable in the traditional scientific framework. 
Their privacy and ineffability are notorious. Qualia are {\em structured}: some of them are more fundamental than others, e.g. spectrum colours, sound pitch, the five different kinds of taste buds (sweet, salty, sour, bitter, and umami), somatic sensations (pain, pleasure, ...),  basic emotions (sadness, happiness, ...). Qualia are compositional with internal structures that generally partially determine their qualitative nature. They are connected by different relations, forming numerous complex structures, such as thoughts and emotions. 
Since we regard consciousness as the direct fruition of a very structured kind of quantum information (we are the system supporting such final information), then, according to principle P1, qualia correspond to information states ``felt'' by the aware system, and according to principles P2 and P2' such states are quantum states of such systems. 
\begin{itemize}
\item[S1:]\hskip4pt Qualia are described by ontic pure quantum states. 
\end{itemize}
We will see in Subsect. \ref{qqualia} how quantum entanglement can account for organizing nontrivial qualia.
The {\em qualia space} thus would be the Hilbert space of the systems involved in the quale. 

Classical information coming from the environment through the senses is ultimately converted into quantum information which is experienced as qualia. The inverse quantum-to-classical transformation is also crucial in converting structures within qualia into logical and geometric representations expressible with classical information to communicate free-will choices to classical structures. For such purposes, the interaction with the memory of past experiences may be essential, for, as we shall see, memory is classical.

\medskip

Finally we come to the problem of {\em free will}. First we must specify that free will, contrary to consciousness, produces public effects which are classical manifestations of quantum information. Since the manifestation is classical, the possible choices of the conscious agent are in principle {\em jointly perfectly discriminable}.\footnote{In the usual classical information theory the convex set of states is a simplex, and its extremal states are jointly discriminable.} 

We then state what we mean by {\em free}.
\begin{itemize}
\item[S2:]\hskip4pt Will is {\em free} if its unpredictability by an external observer cannot be interpreted in terms of lack of knowledge.\footnote{Here there is a clear distinction between the "willing agent" and the observers of it. The word {\em unpredictability} applies only to the observers.}
\end{itemize}
In particular, statement S2 implies that the actual choice of the entity exercising its free will cannot be in principle predicted with certainty by any external observer. One can immediately recognise that if one regards the choice as a random variable, it cannot be a classical one, which can always be interpreted as lack of knowledge. On the other hand, it fits perfectly the case of quantum randomness, which cannot be interpreted as such.\footnote{The impossibility of interpretation of quantum randomness as lack on knowledge fits quantum complementarity. Indeed we cannot by any means know both values of two complementary variables. One may argue that both values may exist anyway, even if we cannot know them, but such argument disagrees with the violation of the CHSH bound (the most popular Bell-like bound), which is purely probabilistic, and is based on the assumption of the existence of a joint probability distribution for the values of complementary observables, assumption that obviously is violated by quantum complementarity. Others can argue (and this is the most popular argument) that one must use different local random variables depending on remote settings, which leads to the interpretation of the CHSH correlations in terms of {\em nonlocality}:
such interpretation, however, is artificial, whereas the most natural one iw that {\em the measurement outcome is created by the measurement}. 
} 

\section{Qualia: the role of entanglement}\label{qqualia}
\begin{figure}[h]
\begin{center}
\begin{tabular}{rl}
\begin{tabular}{r}\includegraphics[width=6cm]{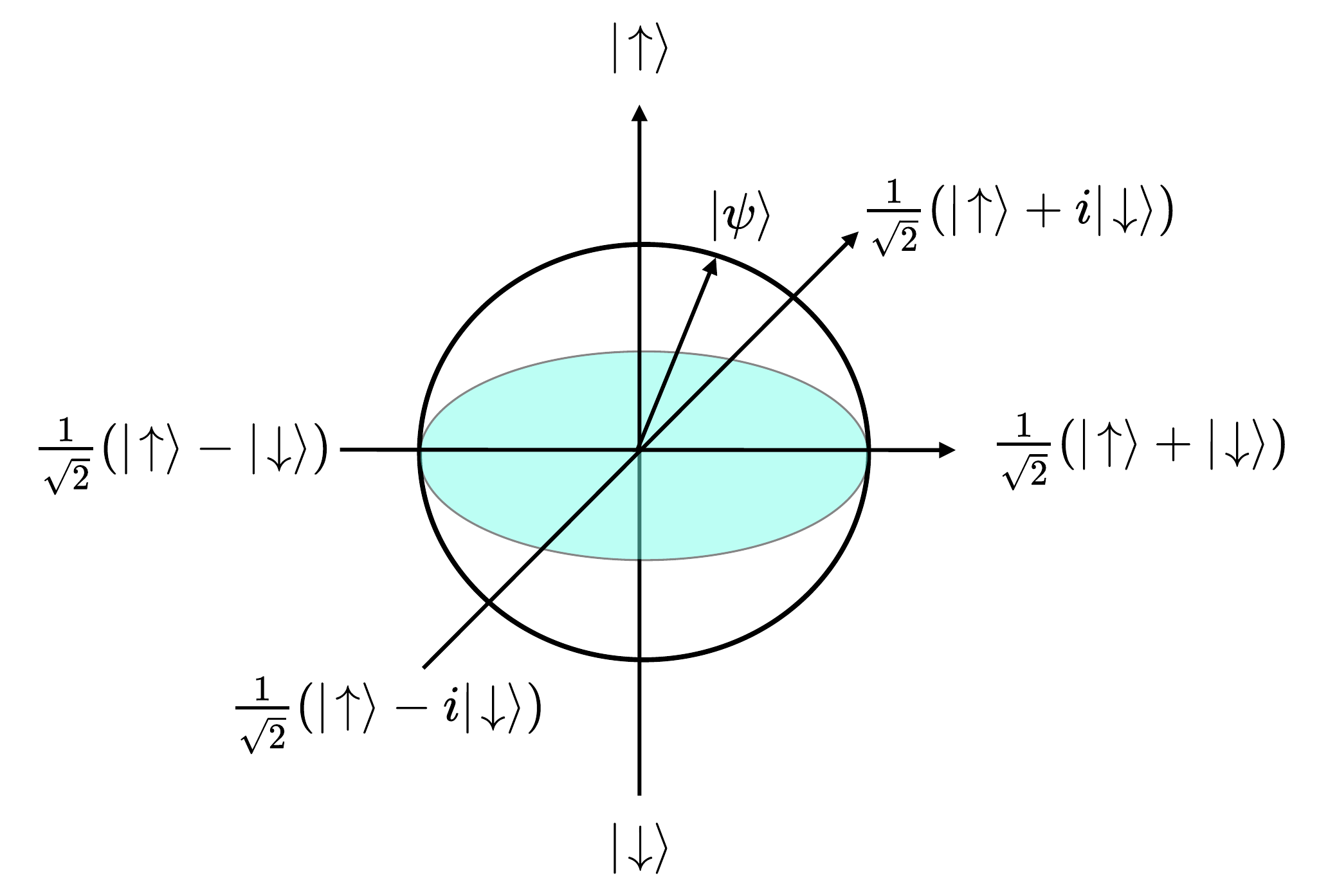}\end{tabular}&
\begin{tabular}{l}\includegraphics[width=5.3cm]{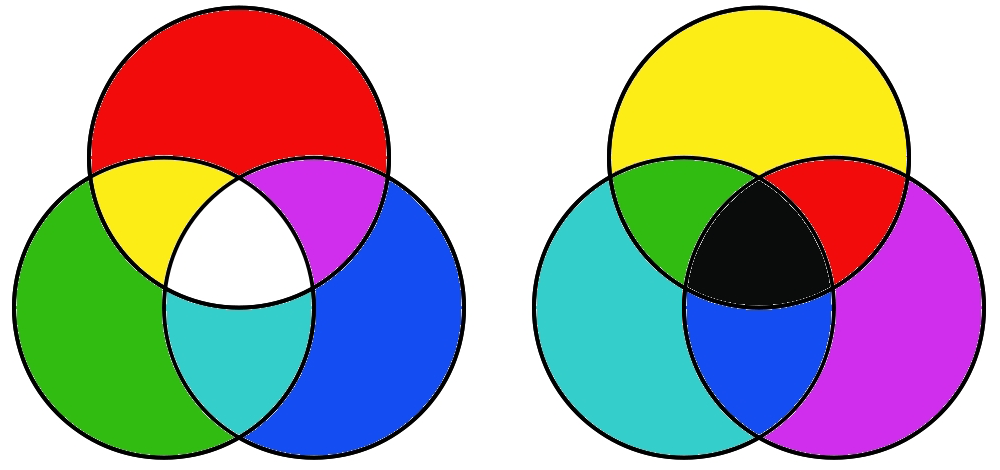}\end{tabular}
\end{tabular}
\caption{\label{fig:flowchart} Illustration of how qualia superpose to make new qualia. {\bf Left figure:} the Bloch sphere illustrates the case of {\em direction} qualia. Summing and subtracting the quale ``up'' with the quale ``down'' we obtain the qualia ``right'' and ``left'', respectively. Similarly, if we make the same combinations with the imaginary unit $i$ in front of  ``down''  we obtain the qualia ``front'' and ``back''. Summing with  generic complex amplitudes $a$ and $b$ (with $|a|^2+|b|^2=1$) we get the quale corresponding to a generic direction,  corresponding to a generic pure state $|\psi\>$. {\bf Right figure:} summations of two  of the three colour basis {\em red, green},  and {\em blue} (RGB) and subtraction of two of the three colours {\em cyan, magenta} and {\em yellow} (CMY). As explained in the main text, these  examples are only for the sake of illustration of the notion of linear combination of quantum ontic states to make new ontic states corresponding to new qualia. A third case of superposition is that of  sounds with precise frequency that combine with addition and subtraction into timbers and chords. Notice that all three cases fit the wave aspect of reality, not the particle one.}\label{f:qualia}
\end{center}
\end{figure}

According to Statement S1 qualia are described by {\em ontic} quantum states and, being such states {\em pure}, we can represent them by normalised vectors $|\psi\>\in\sH_\rA$ in the Hilbert space $\sH_\rA$ of system $\rA$. For the sake of illustration we consider two simple cases of qualia: {\em direction} and {\em colour}. These examples should not be taken literally,\footnote{We emphasize that the above examples are only for the sake of illustration of concepts. The example about color qualia in Fig. \ref{f:qualia} would be a faithful one if the colours were monochromatic and the summation or subtraction were made with wave amplitude, not intensity, which is the actual case. The case of direction qualia based on the Bloch sphere is literally correct. The directions are those of the state representations on the Block sphere.} but only for the sake of illustration of the concept that the linear combinations of qualia can give rise to new complex qualia.

We have said that qualia combine to make new qualia, and {\em thoughts} and {\em emotions} themselves are structured qualia. Superimposing two different kinds of qualia in an entangled way produces new kinds of qualia. Indeed, consider the superposition of a red up-arrow with a green down-arrow. This is not a yellow right-arrow as one may expect, since the latter corresponds to the independent superposition of direction and colour, as in the following equation:
\vskip -10pt
\begin{equation}
\begin{aligned}
\includegraphics[width=10cm]{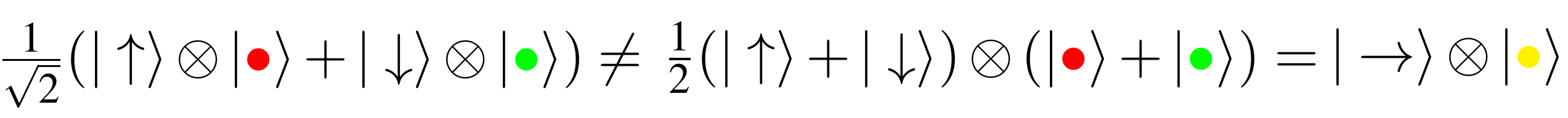}\\[-9pt]
\end{aligned}
\end{equation}
\vskip 6pt\noindent We would have instead
\begin{equation}
\begin{aligned}
\includegraphics[width=7.5cm]{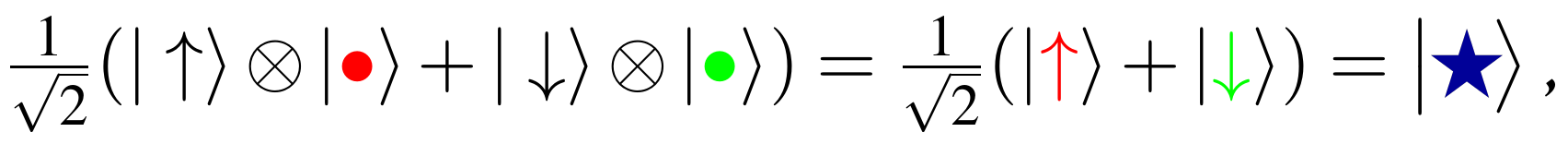}\\[-9pt]
\end{aligned}
\end{equation}
where the ket with the blue star represents a completely new qualia. In a more general case, we have a state vector with triple or quadruple or more entanglement in a factor of a tensor product, and more generally, every system is entangled, e.g. 
\begin{equation}
\ldots \otimes(|a_1\>\otimes|b_1\>\otimes|c_1\>\otimes|d_1\>\pm|a_2\>\otimes|b_2\>\otimes|c_2\>\otimes|d_2\>)\otimes
(|\uparrow\>\otimes|\redb\>\pm|\downarrow\>\otimes|\greenb\>)\otimes\ldots
\end{equation}
We can realise how in this fashion one can achieve new kinds of qualia whose number grows exponentially with the number of systems. In fact, the number of different ways of entangling $N$ systems corresponds to the number of partitions of $N$ into integers, multiplied by the number of permutations of the systems, and therefore it grows as $N!e^{\sqrt{2N/3}}$, and this without considering the variable vectors that can be entangled! 

Since qualia correspond to pure states of the conscious systems, their Hilbert space coincides with the multidimensional Hilbert space of the system. Experimentally one may be able to locate the systems in terms of  neural patterns, but will never be able to read the encoded information without destroying the person's experience, while at best gaining only a single complementary side of the qualia, out of  exponentially many. In synthesis: 
\begin{itemize}
\item System identification is possible, but not the "experience" within.
\end{itemize}

The fact that it is possible to identify the information system proves that the identity of the observer/agent is public and is thus correlated with its "sense of self," which is instead private. This is a crucial requirement for a unified theory of consciousness and free will, namely that the observer/agent be identifiable both privately--from within and through qualia--and publicly--from without and through information. Within QT (or post-quantum OPTs) this is possible.

\section{Evolution of Consciousness and Free Will}
Since the quantum state of a conscious system must be pure at all times, the only way to guarantee that the evolving system state remains pure is that the evolution itself is pure (technically it is {\em atomic}, namely its CP-map $\tT=\sum_iT_i\cdot T_i^\dag$ has a single Krauss operator $T_{i_0}$). We remind that both states and effects are also transformations from trivial to non trivial systems and viceversa, respectively. In Table \ref{t:ontic-transformation} we report the theoretical representations of the three kinds of ontic transformations, including the special cases of state and effect.

\renewcommand{\arraystretch}{1.5}
\begin{table}[h]
\begin{tabular}{|r|c|c|c|c|l|}
\hline
 &Circuit symbol & Symbol & Map & Operator & domain-codomain\\
\hline
transformation&
$\Qcircuit @C=1em @R=.7em @! R { &\poloFantasmaCn{\rA}\qw&\gate{\tT}&\poloFantasmaCn{\rB}\qw&\qw}$        
     &$\tT$  &$\tT=T\cdot T^\dag$ &$T$ & $ \Bnd(\sH_\rA\to\sH_\rB)$\\
\hline
Effect & $\Qcircuit @C=1em @R=.7em @! R { &\poloFantasmaCn{\rA}\qw&\measureD{\alpha}}$&$(\alpha|$ &$\Tr[\cdot |\alpha\>\<\alpha|]$&$\<\alpha|$&$\Bnd(\sH_\rA\to\Cmplx)$\\
\hline
State & $\Qcircuit @C=1em @R=.7em @! R {\prepareC{\psi} &\qw\poloFantasmaCn{\rA}&\qw}$ &$|\psi)$&$|\psi\>\<\psi|$&
$|\psi\>$&$\Bnd(\Cmplx\to\sH_\rA)$\\
\hline
\end{tabular}
\caption{Notations for the three kinds of ontic transformations (for the meaning of symbols see Table \ref{tnotat}). For the list of Quantum Theory axioms and its main theorems see Tables \ref{tabminimalQ} and \ref{tabmainthmsQ}.}
\label{t:ontic-transformation}
\end{table}
Let's consider now the general scenario of a conscious composite system in an ontic state $\omega_t$ at time $t$ evolved by the one-step {\em ontic transformation} $\tO^{(t,x_t)}_{F_t}$ with outcome $F_t$, depending on classical input $x_t$ from senses and memory\footnote{As we will see, memory is classical.} 
\begin{align}
\quad\Qcircuit @C=1em @R=.7em @! R {\prepareC{\omega_t}
     &\qw\poloFantasmaCn{\rA_t}&\gate{\tO^{(t,x_t)}_{F_t}}&\qw\poloFantasmaCn{\;\;\;\rA_{t+1}}&\qw&\qw}\;\;\;=:\;\;\;
    \Qcircuit @C=1em @R=.7em @! R {\prepareC{\omega_{t+1}}
     &\qw&\qw\poloFantasmaCn{\;\;\rA_{t+1}}&\qw&\qw}. 
\end{align}
The {\em epistemic transformation} would be the sum of all ontic ones corresponding to all possible outcomes:
\begin{equation}
\tE^{(t,x_t)}:=\sum_{F_t}\tO^{(t,x_t)}_{F_t}.
\end{equation}
\begin{itemize}
\item[] The outcome $F_t$ is a classical output, and we identify it with the {\em free will} of the experiencing system.
\end{itemize}
 It is a probabilistic outcome that depends on the previous history of qualia of the system. Its kind of randomness is quantum, which means that {\em it cannot be interpreted as lack of knowledge}, and, as such, {\em it is free}. 
Notice that both mathematically and literally the free will is the outcome of a transformation that corresponds to a change of experience of the observer/agent. The information conversion from quantum to classical can also take into account a stage of "knowledge of the will" corresponding to "intention/purpose", namely  "understanding" of which action is taken.

\begin{figure}[h]
\begin{center}
\includegraphics[width=8cm]{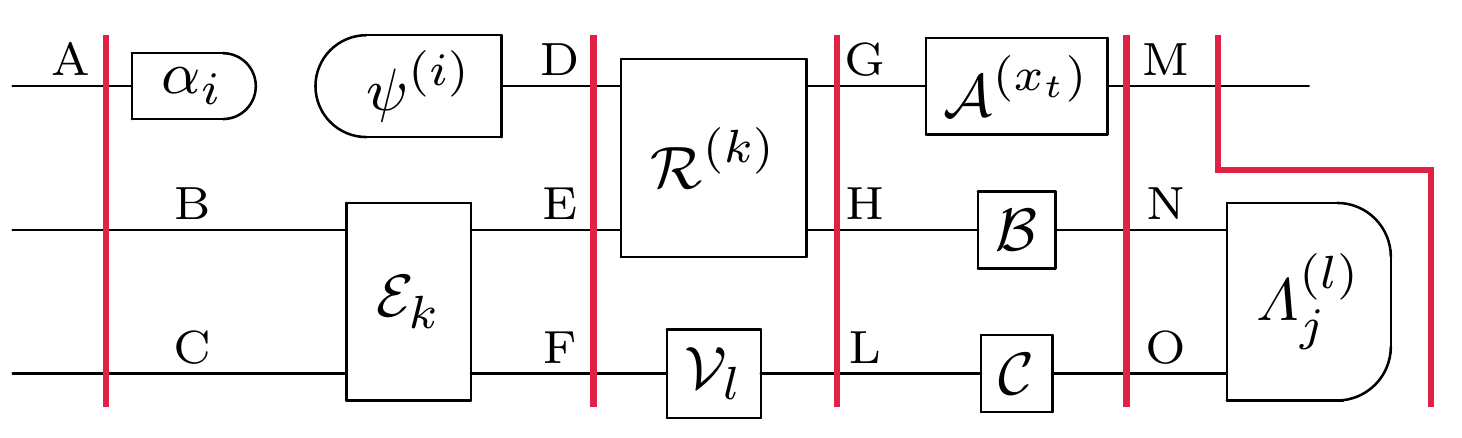}
\caption{Quantum circuit foliation corresponding to Eq. \eqref{Ofoliat}}\label{f:foliation1}
\end{center}
\end{figure}

\smallskip
We may need to provide a more refined representation of the one-step ontic transformation of the evolution in terms of a quantum circuit, for example:

\begin{equation}\label{bigcircuit}
 \Qcircuit @C=1em @R=.7em @! R {&\\
     &\qw\poloFantasmaCn{\rA\rB\rC}&\gate{\tO^{(t,x_t)}_{F_t}}&\qw\poloFantasmaCn{\;\rM}&\qw&\;\;=\;}\:\:\:\:
   \Qcircuit @C=1em @R=.7em @! R {
&\ZS{\rA}		&\Zd{\alpha_i}	&\Zc{\psi^{(i)}}	&\ZS{\rD} 	&\ZG{1}{\tR^{(k)}}  	&\ZS{\rG}&\Zg{\tA^{(x_t)}}&\ZS{\rM} &\qw\\
 &\qw			&\ZS{\rB}	&\ZG{1}{\tE_k}	&\ZS{\rE} 	&\Zh{\tR^{(k)}}		&\ZS{\rH}&\Zg{\tB}&\ZS{\rN} &\ZD{1}{\Lambda_j^{(l)}} \\ 
 &\qw			&\ZS{\rC}	&\Zh{\tE_k}		&\ZS{\rF} 	&\Zg{\tV_l}		&\ZS{\rL}&\Zg{\tC}&\ZS{\rO} &\Zh{\Lambda_j^{(l)}}}\end{equation}\vskip .2cm
\noindent Generally for each time $t$ we have a different circuit.
In the example in circuit \eqref{bigcircuit} we see that a single step can contain also states and effects, and the output system of the whole circuit ($\rM$ in the present case) is generally different (not even isomorphic) to the input ones $\rA\rB\rC$. Following the convention used for the ontic transformation $\tO^{(t,x_t)}_{F_t}$ the lower index is a random outcome,\footnote{
The outcome is random for an observer other than the conscious system, for which, instead, it is precisely known.} and the upper index $(t,x_t)$ is a parameter from which the transformation generally depends. Overall in circuit \eqref{bigcircuit} we have free will $F_t=\{i,k, l,j\}$, whereas the transformations $\psi^{(i)},\tR^{(k)},\tB$, and $\tC$ are deterministic (they have no lower index), and do not contribute to the free will, and the same for the transformation $\tA^{(x_t)}$, which depends on the sensorial input $x_t$. In circuit \eqref{bigcircuit} we can see that {\em we have also classical information at work}, since e.g. the transformation $\tR^{(k)}$ depends on the outcome $k$ of the transformation $\tE_k$, and similarly the choice of test  $\{\Lambda_j^{(l)}\}$ depends on the outcome $l$ of  $\tV_l$, and similarly the state 
$\psi^{(i)}$ depends on the effect outcome $\alpha_i$.
Each element of the circuit is ontic, i.e. atomic, and atomicity of sequential and parallel composition guarantee that the whole transformation is itself atomic. This means that its Krauss operator $O_t$  can be written as the product of the Krauss operators of the component transformations as follows
\begin{equation}
O^{(t,x_t)}_{F_t}=\underbrace{(I_M\otimes \<\Lambda^{(l)}_j|)}_{\rM\rN\rO\to\rM}\underbrace{(A^{(x_t)}\otimes B\otimes C)}_{\rG\rH\rL\to\rM\rN\rO}\underbrace{(R^{(k)}\otimes V_l)}_{\rD\rE\rF\to\rG\rH\rL}\underbrace{(|\psi_i\>\<a_i|\otimes E_k)}_{\rA\rB\rC\to\rD\rE\rF},\label{Ofoliat}
\end{equation}
where $\tO^{(t,x_t)}_{F_t}=O^{(t,x_t)}_{F_t}\cdot O^{(t,x_t)}_{F_t}{}^\dag$.
Notice that in the expression for the operator $O^{(t,x_t)}_{F_t}$ in Eq. \eqref{Ofoliat}  the operators from the input to the output are written from the right to the left--the way we compose operators on Hilbert spaces. Moreover, the expression \eqref{Ofoliat} of $O^{(t,x_t)}_{F_t}$ is not unique, since it depends on the choice of {\em foliation} of the circuit, namely the way you cover all wires with {\em leaves} to divide the circuit into input-output sections. For example, Eq. \eqref{Ofoliat} would correspond to the foliation in Fig. \ref{f:foliation1}

\begin{figure}[h]
\begin{center}
\includegraphics[width=8cm]{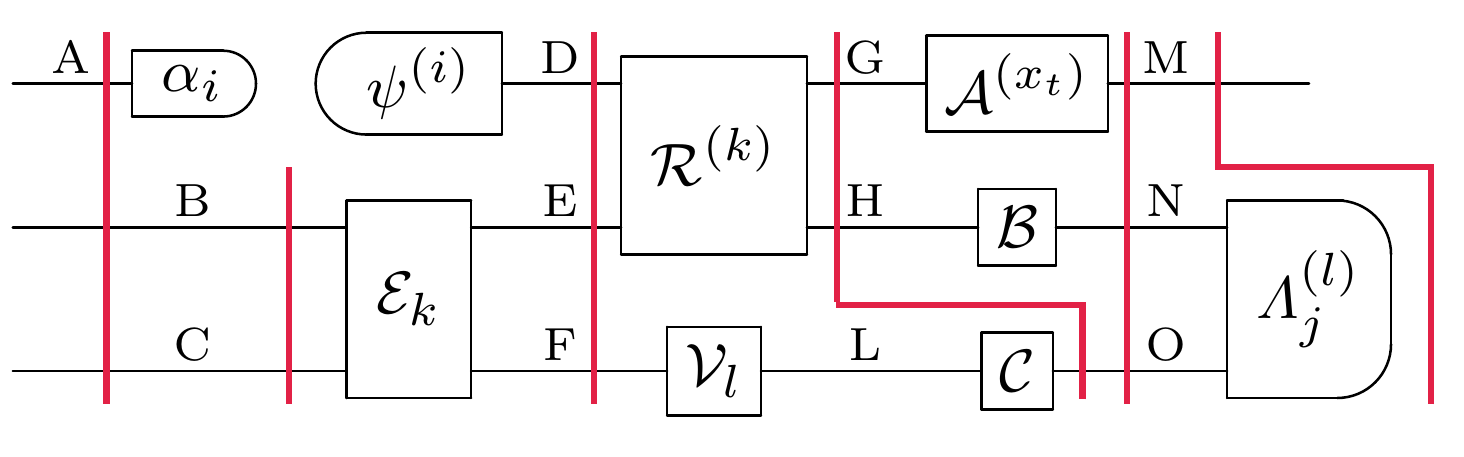}
\caption{Quantum circuit foliation corresponding to the  expression for $O^{(t,x_t)}_{F_t}$ Eq. \eqref{Ofoliat2}}\label{f:foliation2}
\end{center}
\end{figure}

\noindent A different foliation, for example, is the one reported in Fig. \ref{f:foliation2}, corresponding to the expression for $O^{(t,x_t)}_{F_t}$
\begin{equation}
O^{(t,x_t)}_{F_t}=\underbrace{(I_C\otimes \<\Lambda^{(l)}_j|)}_{\rM\rN\rO\to\rM}\underbrace{(A^{(x_t)}\otimes B\otimes I_\rO)}_{\rG\rH\rO\to\rM\rN\rO}\underbrace{(R^{(k)}\otimes CV_l)}_{\rD\rE\rF\to\rG\rH\rO}\underbrace{(|\psi_i\>\otimes E_k)}_{\rB\rC\to\rD\rE\rF}\underbrace{(\<a_i|\otimes I_{\rB\rC})}_{\rA\rB\rC\to\rB\rC}.\label{Ofoliat2}
\end{equation}

\noindent We remind that all operators (including {\em kets} and {\em bras} as operators from and to the trivial Hilbert space $\Cmplx$) are {\em contractions}, namely have norm bounded by 1, corresponding to marginal probabilities not greater than 1. Thus, also $O^{(t,x_t)}_{F_t}$ is itself a contraction. Contractivity for operator $X$ can be conveniently expressed as
\begin{equation}
 X\in\Bnd(\sH_{in}\to\sH_{out}),\quad X^\dag X\leq P_{S_X}\leq I_{in},\quad P_{S_X} :=\Proj\Supp X.
\end{equation}
Now, since the evolution of consciousness must be atomic at all times, we can write a whole consciousness history as the product of the Krauss operators $O_t\equiv O^{(t,x_t)}_{F_t}$ at all previous times $t$
\begin{equation}
\Omega_t:=O_tO_{t-1}\ldots O_1, \quad O_t:= O_{F_t}^{(t,x_t)},
\end{equation}
and apply the history operator to the wavevector of the initial ontic state-vector $|\omega_0\>$
\begin{equation}
|\omega_t\>=\Omega_t|\omega_0\>.
\end{equation}
The squared norm $\n{\omega_t}^2$ of vector $\omega_t$ is the probability of the full {\em history of conscious states} $\{\omega_0,\omega_1,\omega_2,\ldots, \omega_t\}$ and equally of the {\em free-will history} $\{F_1, F_2,\ldots, F_t\}$
\begin{equation}
p(\omega_0,\omega_1,\omega_2,\ldots ,\omega_{t})=p(\emptyset,F_1,\ldots, F_t)=\n{\omega_t}^2=\n{\Omega_t\omega_0}^2.
\end{equation}


\section{Memory}
The ontic evolution of the consciousness state, though it maintains coherence, can keep very limited quantum memory of experience. The latter, due to contractivity of $O$, will go down very fast as $\n{\Omega_t}^2\simeq\n{O}^{-2t}$, i.e. it will decrease exponentially with the number of time-steps. The quantum memory intrinsic in the ontic evolution instead works well as a {\em short-term buffer} to build up a fuller experience--e.g. of a landscape, or of a detailed object, or even to detect motion. How many qubits will make such a single-step buffer? The answer is: not so many.\footnote{For example, the amount of visual information is significantly degraded as it passes from the eye to the visual cortex. Marcus E. Raichle says     \cite{Raichle2010}: 
{\em Of the virtually unlimited information available from the environment only about $10^{10}$ bits/sec are deposited in the retina. Because of a limited number of axons in the optic nerves (approximately 1 million axons in each) only $6*10^6$ bits/sec leave the retina and only $10^4$ make it to layer IV of V1     \cite{Anderson2005,norretranders1998user}.} These data clearly leave the impression that visual cortex receives an impoverished representation of the world, a subject of more than passing interest to those interested in the processing of visual information     \cite{Olshausen2005}. Parenthetically, it should be noted that estimates of the bandwidth of conscious awareness itself (i.e. what we "see") are in the range of 100 bits/sec or less    \cite{Anderson2005,norretranders1998user}.} Indeed, if we open our eyes for  just a second to look at an unknown scene, and thereafter we are asked to answer binary questions, we would get only a dozen of right answers better than chance. Tor N\o rretranders writes     \cite{norretranders1998user}: {\em The bandwidth of consciousness is far lower than the bandwidth of our sensory perceptors.} ... {\em Consciousness consists of discarded information far more than information present. There is hardly information left in our consciousness.}\footnote{We believe that the inability to recall much information contained in one second of visual experience, when the actual experience is felt to be quite rich, should not be construed to diminish the importance of consciousness. In fact, experience is quantum while memory is classical, and although not much classical information appears to have been memorized, the actual experience has the cardinality of the continuum in Hilbert space. Consciousness is about living the experience in its unfolding and understanding what is happening, so to make the appropriate free-will decisions when necessary. Recalling specific information in full detail is unnecessary. Consciousness is focused on the crucial task of getting the relevant meaning contained in the flow of experience. The scarce conscious memory of specific objects and relationships between objects should not be an indication that consciousness is a "low bandwidth" phenomenon, but that what is relevant to consciousness may not be what the interrogator believes should be relevant. }

\medskip
We reasonably deduce that there is actually no room for long-term memory in consciousness, and we conclude that:
\begin{itemize}
\item[S3:]\hskip4pt Memory is classical. Only the short-term buffer to collect each experience is quantum.
\end{itemize}
By ``short'' we mean comparable to the time in which we collect the full experience, namely of the order of a second.

\subsubsection*{Transferring quantum experience to classical memory}

If we have quantum experiences and classical memory, we need to convert information from quantum to classical in the memorization process, and from classical to quantum in the recollection process.  The first process, namely transferring quantum experience to classical memory, must be necessarily incomplete, otherwise it would violate the quantum no-cloning theorem. The Holevo theorem    \cite{Holevo1973BoundsFT} establishes that the maximal amount of classical information that can be extracted from a quantum system is a number of bits equal to the number of qubits that constitute the quantum system. Obviously, such maximal classical information is infinitesimal compared to the continuum of classical  information needed to communicate classically a quantum state! 

From a single measurement one can extract classical information about just one of the continuum of complementary aspects of the quantum state, e.g. along a given direction for a spin. Moreover, the state after the measurement would be necessarily disturbed\footnote{For example, for the spin originally oriented horizontally and measured vertically {\em a la} von Neumann, the final state would be vertical up or down, depending on the measurement outcome.}, due to the {\em information-disturbance tradeoff }.\footnote{There are many ways of regarding the information-disturbance tradeoff, depending on the specific context and the resulting definitions of {\em information} and {\em disturbance}. The present case of  {\em atomic measurements} with the "disturbance" defined in terms of the {\em probability of reversing the measurement transformation} has been analysed in the first part of Ref.     \cite{d2003heisenberg}. In the same reference it is also shown that a reversal of the measurement would provide a contradicting information which numerically cancels the information achieved from the original measurement, thus respecting the quantum principle of no information without disturbance.}  So, which measurement should one perform for collecting the best classical information from a quantum system? 

The answer is easy for a single qubit. Just perform the usual von Neumann measurement along a random direction! This has been proved in Ref. \cite{chiribella2007continuous}.\footnote{Let's consider the case of a single qubit realised with a particle spin. The usual observable of von Neumann corresponds to a measurement of the orientation of the spin along a given direction, e.g. {\em up} or {\em down} along the vertical, or {\em left} or {\em right} along the horizontal direction. But when we prepare a spin state (e.g. by Rabi techniques), we can put the spin in a very precise direction, e.g. pointing north-east along the diagonal from south-west to north-east, and, indeed if we measure the spin along a parallel direction we find the spin always pointing north-east! So the spin is indeed (ontically!) pointing north-east along the same diagonal! Now it would be a legitimate question to ask: {\em how about measuring the direction itself of the spin?} This can be done--not exactly, but optimally--using a continuous observation test constituted of a continuous of effects (what is usually called POVM, the acronym of Positive Operator Valued Measure). However, it turns out that the measured direction is a fake, since such a quantum measurement with a continuous set of outcomes is realised as a continuous random choice of a von Neumann measurement    \cite{chiribella2007continuous}. In conclusion the optimal measurement of the spin direction is realised by a customary Stern-Gerlach experiment in which the magnetic field is randomly oriented! The same method can be used to achieve an informationally complete measurement to perform a quantum tomography  \cite{DAriano2000} of the state, by suitably processing the outcome depending on the orientation of the spin measurement.}
For larger dimensions it is in principle possible to generalise this result, however, the probability space turns out to be geometrically much more complicated than a sphere,\footnote{The convex set of deterministic states for a {\em qutrit} (i.e. $d=3$), defined by algebraic inequalities, has eight dimensions, and has a boundary made of a continuous of balls.} and a  scheme for random choice is unknown. For this reason we can consider the larger class of observation tests called {\em informationally complete} ("infocomplete"),\footnote{What in OPT language is called observation test is the same of what in the quantum information literature is called discrete POVM.
An observation test is infocomplete for system $\rA$ when it spans the linear space of effects $\Eff_\Reals(\rA)$.} which, unlike the case of the optimal measurement of spin direction, can be taken as discrete and with a finite number of outcomes. Such a measurement would correspond to the following epistemic transformation 
\begin{equation}\label{Epistemicinfocomp}
\tM\in\Trn_1(\rA\to\rB),\qquad \tM=\sum_{j\in J} M_j\rho M_j^\dag,\quad |J|\geq \dim(\sH_\rA)^2,
\end{equation}
which is the coarse graining of the {\em infocomplete} quantum test $\{\tM_j\}_{j\in J}$, with $\tM_j= M_j\cdot M_j^\dag$ where $\{|M|^2_j\}_{j\in J}$ is an {\em infocomplete observation test}. 
"Infocomplete" means that, in the limit of infinitely many usages over the same reprepared state, the measurement allows to perform a full tomography  \cite{DAriano2000} of the state.

The infocomplete measurement (or the random observables seen before) is better suited to extract classical information from the quantum buffer for long-term memory, since is does not privilege a particular observable. It is also likely that the conscious act of memorising an experience may be achieved by actually {\em repreparing the ontic state multiple times in the quantum buffer}, and performing the infocomplete test multiple times, thus with the possibility of memorising a (possibly partial) tomography of the state. Notice that, generally, the ontic state $M_j\rho M_j^\dag$ for a given $j$ still depends on $\rho$. However, the test will disturb it, and the less disturbance, the smaller the amount of classical information that can be extracted    \cite{d2003heisenberg}.\footnote{Particularly symmetric types of  measurements are those made with a SIC~POVM (Symmetric informationally complete POVM)    \cite{Fuchs2017}, where $M_j=|\psi_j\>\<\psi_j|$ are $d^2$ projectors on pure states with equal pairwise fidelity
\begin{equation} \label{eq:SICinner}
|\langle\psi_j|\psi_k\rangle|^2=\frac{d\delta_{jk}+1}{d+1}.
\end{equation}
The projectors $|\psi_j\>\<\psi_j|$ defining the SIC POVM in dimension~$d$ form a $(d^2-1)$-dimensional regular simplex in the space of Hermitian operators. }


\subsubsection*{Transferring classical memory to quantum experience}
We have seen that an infocomplete measurement is needed to (approximately) store an experience in the classical long-term memory. By definition, the recovery of an experience requires a reproduction of it, meaning that the corresponding ontic state is (approximately) reprepared from the classical stored data. 
The memory, being classical, will be read without disturbance, thus left available to following recollections. In order to transfer classical to quantum information, methods of state-preparation have been proposed in terms of quantum circuit schemes    \cite{PhysRevA.83.032302}. 

A possible benchmark for the  memory store-and-recall  process is the maximal fidelity achievable in principle with a {\em measure-and-reprepare} scheme that optimises the fidelity between the experienced state and the recalled one. We will consider $M\geq 1$  input copies of the same state to quantify {\em attention} to the experience. The optimal {\em fidelity}\footnote{The fidelity $F$ between two pure states corresponding to state vectors $|\psi\>$ and $|\varphi\>$, respectively, is defined as $|\<\varphi|\psi\>|^2$.} between the experienced state and the recalled one averaged over all possible experiences (i.e. input states) is given by \cite{PhysRevA.72.032325} 
\begin{equation}\label{Md}
F(A,M,d)=\frac{M+1}{M+d},       
\end{equation}
where $d$ is the dimension of the experiencing system.\footnote{The optimal fidelity in Eq. \eqref{Md} is achieved by an observation test with atomic effects $|\Phi_i\>\<\Phi_i|$ with 
\begin{equation}
|\Phi_i\>=\sqrt{w_i d_M}{|\psi_i\>^{\otimes M}},\qquad d_M={{M+d-1}\choose{d-1}},
\end{equation}
where $\{|\psi_i\>\}_{ i=1}^A$ are pre-specified pure state-vectors, and $\{w_i\}_{ i=1}^A$ is a probability vector satisfying the identity
\begin{equation}
\sum_{i=1}^A w_i|\psi_i\>\<\psi_i|^{\otimes M}=\Big\<|\psi\>\<\psi|^{\otimes M}\Big\>.
\end{equation} 
where the notation $\<\ldots\>$ denotes averaging over the prior of pure-state vectors, taking as the prior the Haar measure on the unit sphere in $\Cmplx^d$.
}\par\noindent Such upper bound for fidelity may be used to infer an effective dimension $d$ of the system involved in consciousness, e.g. in very focused restricted experiences as those involved in masking conscious perception   \cite{dehaene2014consciousness}.


\subsection*{Information transfer from body to consciousness and vice versa}
We would expect that most of the operation of the human body is automatic and uses classical information that is never translated into quantum information to be experienced by consciousness. The portion of the classical information produced by the body that needs to be converted into quantum should only be the salient information that supports the qualia perception and comprehension necessary to "live life" and make appropriate free-will choices. The amount of quantum information to be translated into classical for the purpose of free-will control of the body top-down should be relatively small.

\section{On the combination problem}\label{s:causal}
The {\em combination problem} concerns the issue about {\em if and how the fundamental conscious minds come to compose, constitute, or give rise to some other, additional conscious mind}    \cite{Chalmers2016-CHATCP-6}.  By definition, the problem becomes crucial for panpsychism: if cosciousness is everywhere, what is the criterion to select novel conscious individuals? Is the union of two conscious beings a conscious being? If this is true, then any subset of a conscious being can also be a conscious being. The present theoretical approach provides a precise individuation criterion. The criterion derives from principle P3 about the purity of quantum conscious states and, consequently, the need for ontic transformations. Let's see how it works.

\medskip
It is reasonable to say that {\em an individual is defined by the continuity of its experience}. 
Such a statement may be immediately obvious to some readers. However, for those who may not agree, we propose a thought experiment.

\smallskip
{\em Consider a futuristic ``quantum teleportation experiment'', meaning that the quantum state of a system is substituted to that of a remote isomorphic system.\footnote{Teleportation will need the availability of shared entanglement and classical communication, and technically would use a Bell measurement at the sender and a conditioned unitary transformation at the receiver.    \cite{tele}} The matter (electrons, protons, ...) of which a teleported person is composed is available at the receiver location and is in-principle indistinguishable from the same matter at the transmission location: indeed, it is only its quantum state that is teleported. The resulting individual would be the same as the original one, including his thoughts and memory.}\footnote{Of course, this would violate the no cloning theorem, if at the transmission point the original individual would not be destroyed. Indeed, according to the quantum information-disturbance tradeoff     \cite{d2003heisenberg}, teleportation cannot even make a bad copy leaving the original untouched.}
\medskip

\begin{figure}[h]
\begin{center}
\includegraphics[width=11.5cm]{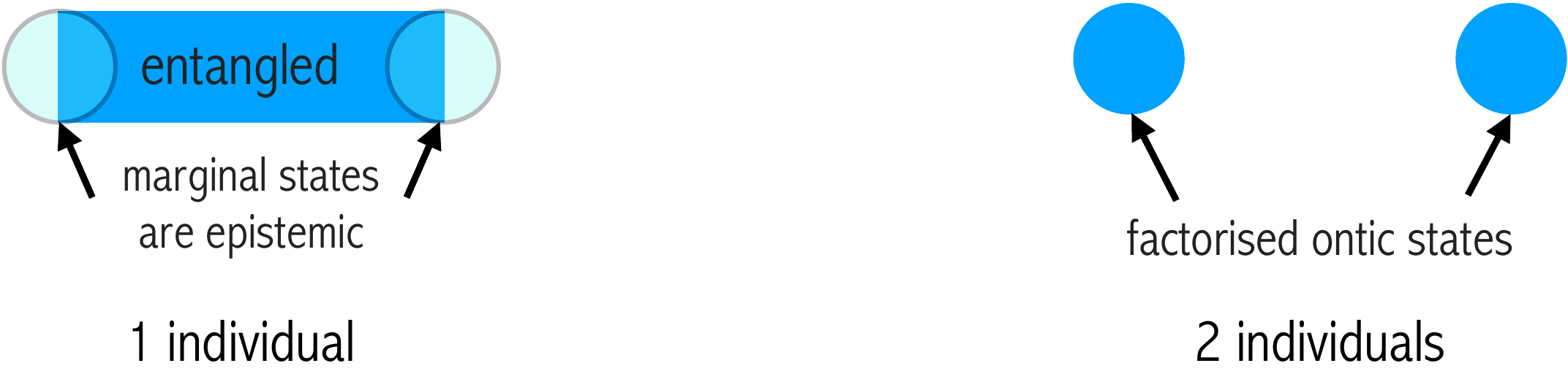}
\caption{{\bf The combination problem.} 
Stated generally, the problem is about how the fundamental conscious minds come to compose, constitute, or give rise to some further conscious mind. The ontic-state principle P1 provides a partial criterion to exclude some situations, e.g. (figure on the left) two entangled systems cannot separately be  conscious entities, since each one is in a marginal state of an entangled one, hence it is necessarily mixed. On the other hand, (figure on the right) if two systems are in a factorized pure state, each system is in an ontic state, and they are two single individuals, and remain so depending on their following interactions (see Fig. \ref{f:conscious-minds}). }\label{f:comb-problem}
\end{center}
\end{figure}

\begin{figure}[h]
\begin{center}
\includegraphics[width=11.5cm]{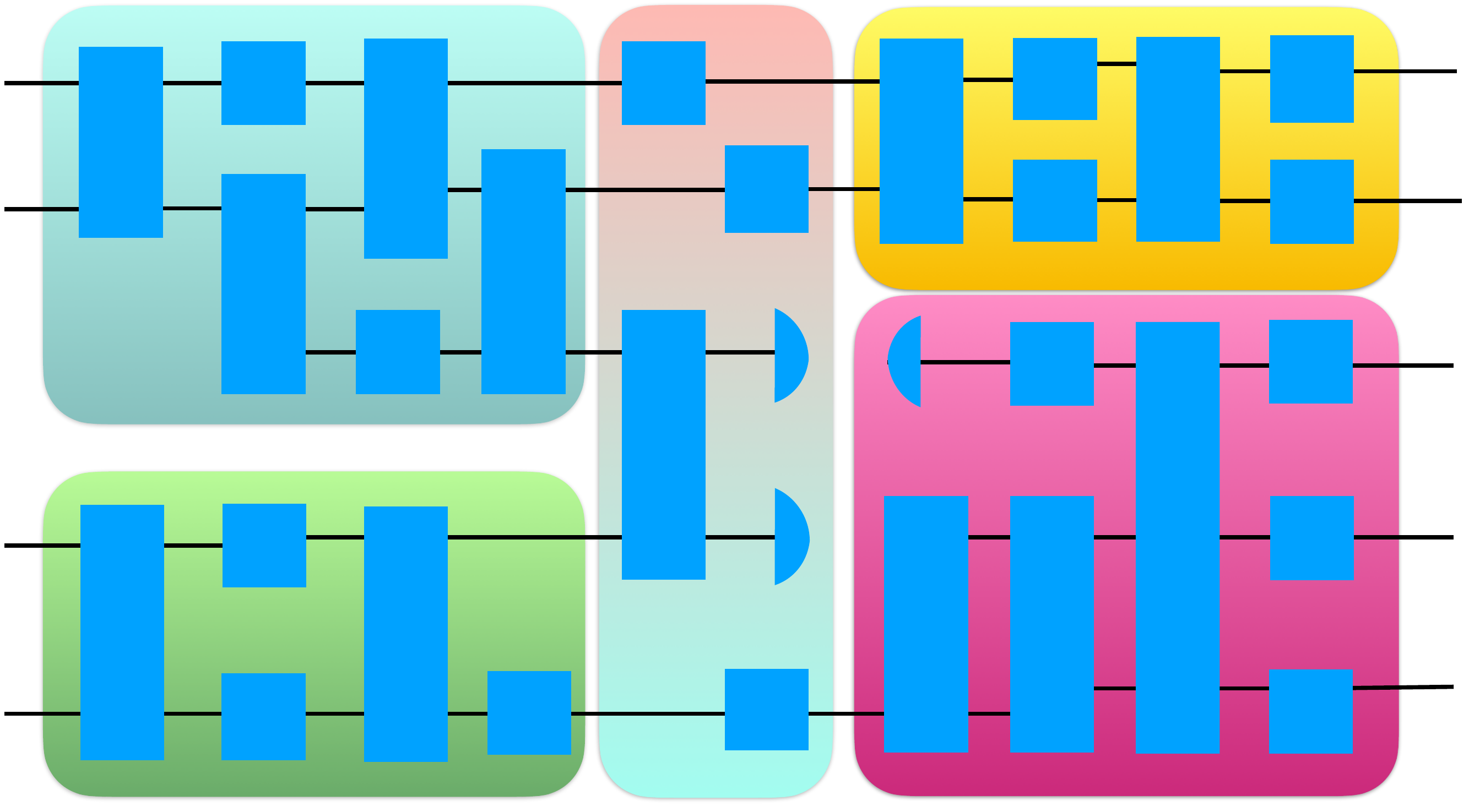}
\caption{{\bf The combination problem.} 
The general individuation criterion in statement S4 requires full quantum coherence, namely the ontic nature of both states and transformations. In the case depicted in this figure, every transformation (including effects and states as special cases) is pure, and we suppose each multipartite transformation is not factorizable. Then the first two boxes on the left  represent two separate individuals. The box in the middle merges the two individual into a single one, whereas the immediately following effects convert quantum to classical information, and separate again the single individual into the original separate ones. Notice that a merging of two individuals necessarily needs a quantum interaction.} \label{f:conscious-minds}
\end{center}
\end{figure}

The above thought experiment suggests the following {\em individuation criterion} within our theoretical approach:
\begin{itemize}
\item[S4:]\hskip4pt A conscious mind is a composite system in an ontic state undergoing an ontic transformation,  with no subsystem as such. 
\end{itemize}
In Figs. \ref{f:comb-problem} and \ref{f:conscious-minds} we illustrate the use of the criterion in two paradigmatic cases. Here we emphasise that the role of interactions is crucial for the criterion.

This section has described what is necessary to form a combination of conscious entities, thus removing the difficulties encountered with panpsychist models based on classical physics. The key idea is that the states of the combining systems and their transformations be ontic and that such systems interact quantumly. These requirements assure that the combined entity is also in a pure state and that {\em none of its subsets are conscious}. Clearly, the interactions between the conscious entity and the environment (including other entities) can only be classical, otherwise a larger entity would be created and would persist for as long as a disentangling transformation does not occur, as in the case illustrated in Fig. \ref{f:conscious-minds}.

\section{Theory experimentability and simulability}
Proposing feasible experiments about the quantum nature of consciousness is a very exciting challenge. In principle quantumness could be established through experimental demonstration of one of the two main nonclassical features of quantum theory: {\em nonlocality} and {\em complementarity}. The two notions are not independent, since in order to prove nonlocality we need complementary observations, in addition to shared entanglement. 

In order to prove nonlocality of consciousness we need measurements at two separate points sufficiently far apart to exclude causal connection, and such a requirement is very challenging, since it demands very fast measurements, considering that for a distance of 3 cm  between the measurements points a time-difference of a tenth of nanosecond is sufficient to have signalling. 

About nonlocality together with complementarity, we speculate that they together may be involved in 3D vision, and take the opportunity to suggest that such line of research may anyway be a fruitful field of experimental research on consciousness.\footnote{A paradigmatic case of superposition between incompatible 3D views is that of the Necker cube, which some authors regard as neuro-physiological transformation leading to perceptual reversal controlled by the principles of quantum mechanics    \cite{Caglioti-Benedek-Neckercube}. However, the 3D experience of the Necker cube does not require binocular disparity since monocular vision also produces the same effect. 
} 
For example, a genuine case of complementarity in 3D experience is occurring as a result of looking at {\em Magic Eye} images published in a series of books     \cite{Magiceye}. These images feature {\em autostereograms}, which allow most people to see 3D images by focusing on 2D patterns that seem to have nothing in common with the 3D image that emerges from them. The viewer must diverge or converge his eyes in order to see a hidden three-dimensional image within the patterns. The 3D image that shows up in the experience is like a glassy object that, depending on convergence or  divergence of the eyes, shows up as either concave or convex. Clearly the two alternative 3D views--convex and concave--are truly complementary experiences, and each experience has an intrinsic wholeness. 
 
Regarding complementarity alone, speculative connections with contrasting  or opposite dimensions of human experience have been considered in the literature, e.g. "analysis" vs "synthesis", and "logic" vs "intuition"    \cite{Jahn2007}. Here, however, we are interested in {\em complementariy in experiences possibly reproducible in different individuals}, of the kind of  {\em gedanken-experiments a la Heisenberg}, e.g. one experience incompatible with another and/or disturbing each other. Having something more than two complementary observables, namely an informationally complete set of measurements such as all three Pauli matrices for a single consciousness qubit,
 would allow us to make a quantum tomography  \cite{DAriano2000} of the qubit state, assuming this is reprepared many times, e.g. through an intense prolonged attention. To our knowledge, the feasibility assessment of such a kind of experiment is of a difficulty comparable to that of a nonlocality experiment.

Finally, apart from proving the quantumness of experience, we can at least experimentally infer some theoretical parameters in the quantum theoretical approach consistent with observations.  This is the case e.g. of the experiment already mentioned in discussing the upper bound \eqref{Md} in memory-recall fidelity, which can be used in inferring an {\em effective dimension} $d$ of the system involved in consciousness in very focused restricted experiences, as for masking conscious perception   \cite{dehaene2014consciousness}. We believe that memory-recall experiments versus  variables such as attention, memory-recovery delay, and variable types of qualia, may be helpful to make a preliminary mapping of the dimensionality of the spaces of conscious systems involved.

\medskip
We conclude with a few considerations about the feasibility of a simulation of a conscious process like the ones proposed here. As already mentioned, due to the purity of the ontic process, a simulation just needs multiplications of generally rectangular matrices, which for matrices with the same dimension $d$ is essentially a $\Theta(d^3)$ process. For sparse matrices (as it is often the case in a quantum simulation) the process can be speeded up considerably. However, to determine the probability distribution of each ontic step, one needs to evaluate the conditional probability for the output state of the previous step, and this needs the multiplication for all possible outcomes (the free will) of the corresponding Kraus operator of the last ontic transformation. With a RAM of the order of Gbytes one could definitely operate with a dozen qubits at a time. This should be compared with 53 qubits of the Google or IBM quantum computer, the largest currently available, likely to share classical information in tandem with a large classical computer. 
However, it is not excluded that some special phenomena could be already discovered/analyzed with a laptop.

\medskip
\section{Philosophic implications}
We have presented a theory of consciousness, based on principles, assumptions, and key concepts that we consider crucial for the robustness of the theory and the removal of the limitations of most panpsychist theories \cite{Chalmers2016-CHATCP-6}. We believe that conferring inner reality and agency to quantum systems in pure quantum state, with conversion of information from classical to quantum and vice versa, is unprecedented, with major philosophical and scientific consequences. In the present approach free will and consciousness go hand in hand, allowing a system to act based on its qualia experience by converting quantum to classical information, and thus giving causal power to subjectivity--something that until now has been highly controversial, if not considered impossible. 

The theory provides that a conscious agent may intentionally convert quantum information into a specific classical information to express its free will, a classical output that is in principle unpredictable due to its quantum origin. The theory would be incoherent without the identification of the conscious system in terms of purity and inseparability of the quantum state, which is identified with the systems experience. The purity of non-deterministic quantum evolution identifies consciousness with agency through its outcome. 

Metaphysically the proposed interpretation that a pure, non-separable quantum state is a state of consciousness, could be turned on its head by assuming the ontology of consciousness and agency as primary, whereas physics is emergent from consciousness and agency. This same interpretation would then consider classical physics the full reification (objectification) of quantum reality	 as quantum-to-classical agency corresponding to the free will of conscious entities existing entirely in the quantum realm. The ontology deriving by the acceptance of consciousness as fundamental, would be that objectivity and classical physics supervene on quantum physics, quantum physics supervenes on quantum information, and quantum information supervenes on consciousness. If we were to accept this speculative view, physics could then be understood as describing an open-ended future not yet existing because the free will choices of the conscious agents have yet to be made. In this perspective, we, as conscious beings, are the co-creators of our physical world. We do so individually and collectively, instant after instant and without realizing it, by our free-will choices.


\appendix
\addcontentsline{toc}{section}{Appendix}
\section*{Appendices about general OPTs}
In these appendices we provide the general operational probabilistic framework
for future post-quantum explorations for a theory of consciousness. We report just a short illustration of  what is an operational probabilistic theory (OPT), of which Quantum Theory and Classical Theory are the most relevant istances. As the reader may appreciate, the OPT provides a framework much reacher, general, flexible, and mathematically rigorous as compared to other theoretical frameworks, such as the causal approach of  Tononi \cite{Tononi2008}.
\subsubsection*{The operational probabilistic theory (OPT) framework}\label{OPTs}
It is not an overstatement to say that the OPT framework represents a new Galilean revolution for the scientific method. In fact, it is the first time that a theory-independent set of rules is established on how to build up a theory in physics and possibly in other sciences. Such rules constitute what is called the {\em operational framework}. Its rigour is established by the simple fact that the OPT is just "metamathematics", since it is a chapter of {\em category theory}     \cite{maclane,selinger2011survey}. To be precise  the largest class of OPTs corresponds to a {\em monoidal braided category}. The fact that the same categoric framework is used in computer science      \cite{coecke,practicingcoecke,coeckerev,kinder,picturialism} gives an idea of the thoroughness and range of applicability of the rules of the OPT.  

Lucien Hardy in several seminal papers    \cite{hardy2001,causaloid,duotenz}
introduced a heuristic framework that can be regarded as a forerunner of the OPT, which made its first appearance in Refs.     \cite{purification,QUIT-Arxiv}, and soon was connected to the categorical approach in computer science     \cite{CoeckePV,Coecke2012,Coecke2016,abramsky_heunen_2016,Coecke2018,TullPV}. As already mentioned, both QT and CT are OPTs     \cite{CUPDCP}, but one can build up other OPTs, such as variations of QT e.g. fermionic QT    \cite{supersel}, or QT on real Hilbert space     \cite{CUPDCP}, or QT with only qubits, but also CT with entanglement    \cite{PhysRevA.101.042118} or without local discriminability      \cite{CTnolocdiscr}. Also other toy-theories, such as the PR-Boxes,  are believed to be completable to OPTs (see e.g.     \cite{barrett}).

The connection of OPT with computer science reflects the spirit of the OPT, which essentially was born on top of the new field of quantum information. Indeed, the OPT framework is the formalization of the rules for building up quantum circuits and for attaching to them a joint probability: in such a way the OPT literally becomes an extension of probability theory. 

\subsubsection*{The general idea}
\begin{figure}[h]
\begin{center}
\begin{tabular}{cc}
\includegraphics[width=11.7cm]{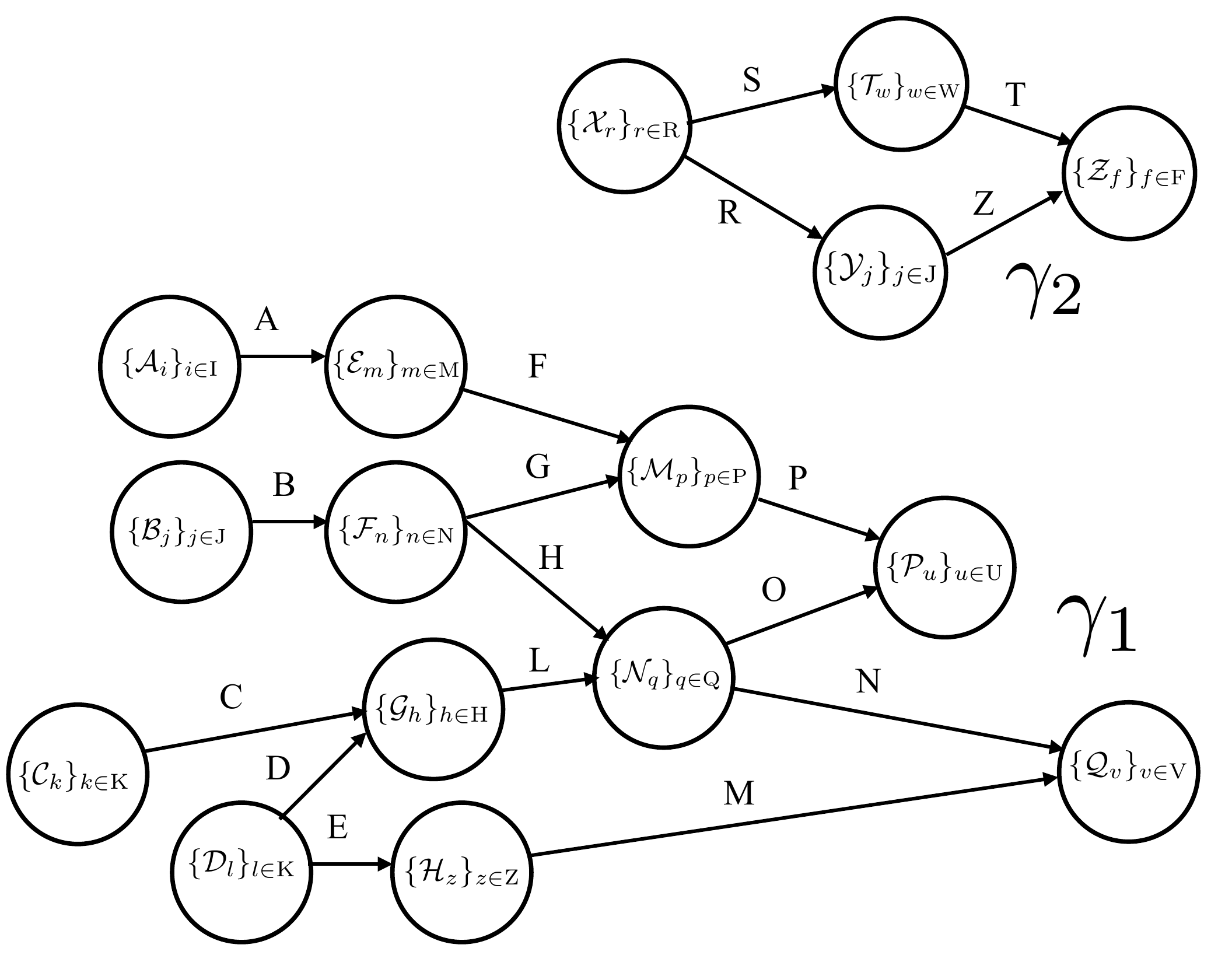} 
\end{tabular}
\caption{\label{fig:flowchart} An OPT associates to each joint probability distribution of multiple tests/events a closed directed acyclic graph (CDAG) of output-input relations (see text). As a rule, unconnected graph components are statistically independent, e.g. in the case in the figure, one has $p(\tA_i,\tE_m,\ldots,\tX_r,\ldots|\gamma_1\cup \gamma_2)=p(\tA_i,\tE_m,\ldots|\gamma_1,)p(\tX_r,\ldots|\gamma_2)$.
}\label{f:DAGOPT}
\end{center}
\end{figure}

How  does an OPT works? It associates to each joint probability of multiple events a {\em closed directed acyclic graph} (CDAG) of input-output relations as in Fig. \ref{f:DAGOPT}. Each {\em event} (e.g. $\tE_m$ in figure) is an element of a complete {\em test} ($\{\tE_m\}_{m\in\rM}$ in figure) with normalized marginal probability $\sum_{m\in\rM}p(\tE_m)=1$. The graph tells us that the marginal probability distribution of any set of  tests still depends on the marginalized set, e.g. the marginal probability distribution of  test $\{\tE_m\}_{m\in\rM}$ depends on the full graph of tests to which it is connected, although it has been partially or fully marginalised. As a rule, disconnected graphs (as $\gamma_1$ and $\gamma_2$ in Fig. \ref{f:DAGOPT}) are statistically independent, namely their probability distributions factorize.
The oriented wires denoting output-input connections between the tests (labeled by Roman letters in figure \ref{f:DAGOPT}), are the so called {\em systems} of the theory. 

\subsubsection*{A paradigmatic quantum example}
\begin{figure}
\begin{center}
\includegraphics[width=8cm]{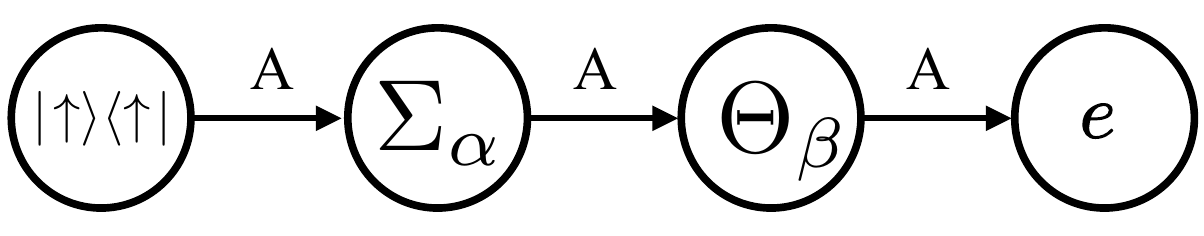} 
\caption{{\bf A simple OPT graph.} A paradigmatic example
for the sake of  illustration and for motivation (see text).}\label{f:parex}
\end{center}
\end{figure}

A paradigmatic example of OPT graph is given in Fig.\ref{f:parex}. There, we have a source of particles all with spin up (namely in the state $\rho=|\uparrow\>\<\uparrow|$). At the output we have two von Neumann measurements\footnote{A von Neumann measurement of e.g. $\sigma_z$ has two outcomes "up" and "down", and the output particle will be in the corresponding eigenstate of $\sigma_z$.} in cascade, the first one $\Sigma_\alpha$ measuring $\sigma_\alpha$, and the second one $\Theta_\beta$ measuring $\sigma_\beta$, where $\alpha$ and $\beta$ can assume either of the two values $x,z$. The setup is  represented by the graph shown in the figure, where $\rA$ represents the system corresponding to the particle spin,  $e$ the deterministic test that simply discards the particle, and the two tests
$\Sigma_\alpha,\Theta_\beta$ ($\alpha,\beta=x,z$) the two von Neuman measurements. Now, clearly, for $\alpha=z$ one has the marginal probability distribution $p(\Sigma_z)=(1,0)$ and $p(\Sigma_x)(\tfrac{1}{2},\tfrac{1}{2})$ independently of the choice of the test $\Theta_\beta$. On the other hand, for the second test one has marginal probability $p(\Theta_x)=(1,0)$ for $\Sigma_z$ and $p(\Theta_x)=(\tfrac{1}{2},\tfrac{1}{2})$ for $\Sigma_x$. We conclude that {\em the marginal probability of $\Sigma_\alpha$ is independent of the choice of  the test $\Theta_\beta$, whereas the  marginal probability of $\Theta_\beta$ depends of the choice of  $\Sigma_\alpha$.}
Thus, the marginal probability of $\Theta_\beta$ generally depends on the choice of  $\Sigma_\alpha$, and this concept goes beyond the content of joint probability, and needs the OPT graph. Theoretically, we conclude that there is "something flying" from test $\Sigma_\alpha$ to test $\Theta_\beta$ (although we cannot see it!): this is what we theoretical describe as a spinning particle! This well illustrates the notion of {\em system}: a theoretical connection between tested events.

\subsubsection*{A black-box approach}
\begin{figure}[h]
\begin{center}
\includegraphics[width=5cm]{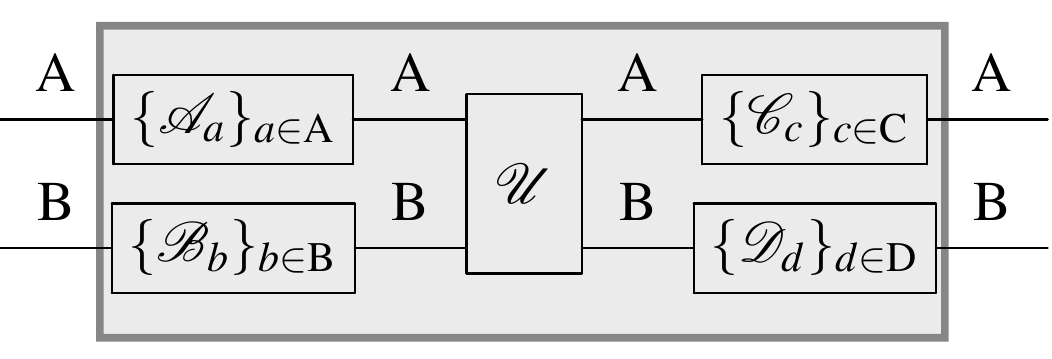}\hskip 1cm
\includegraphics[width=5cm]{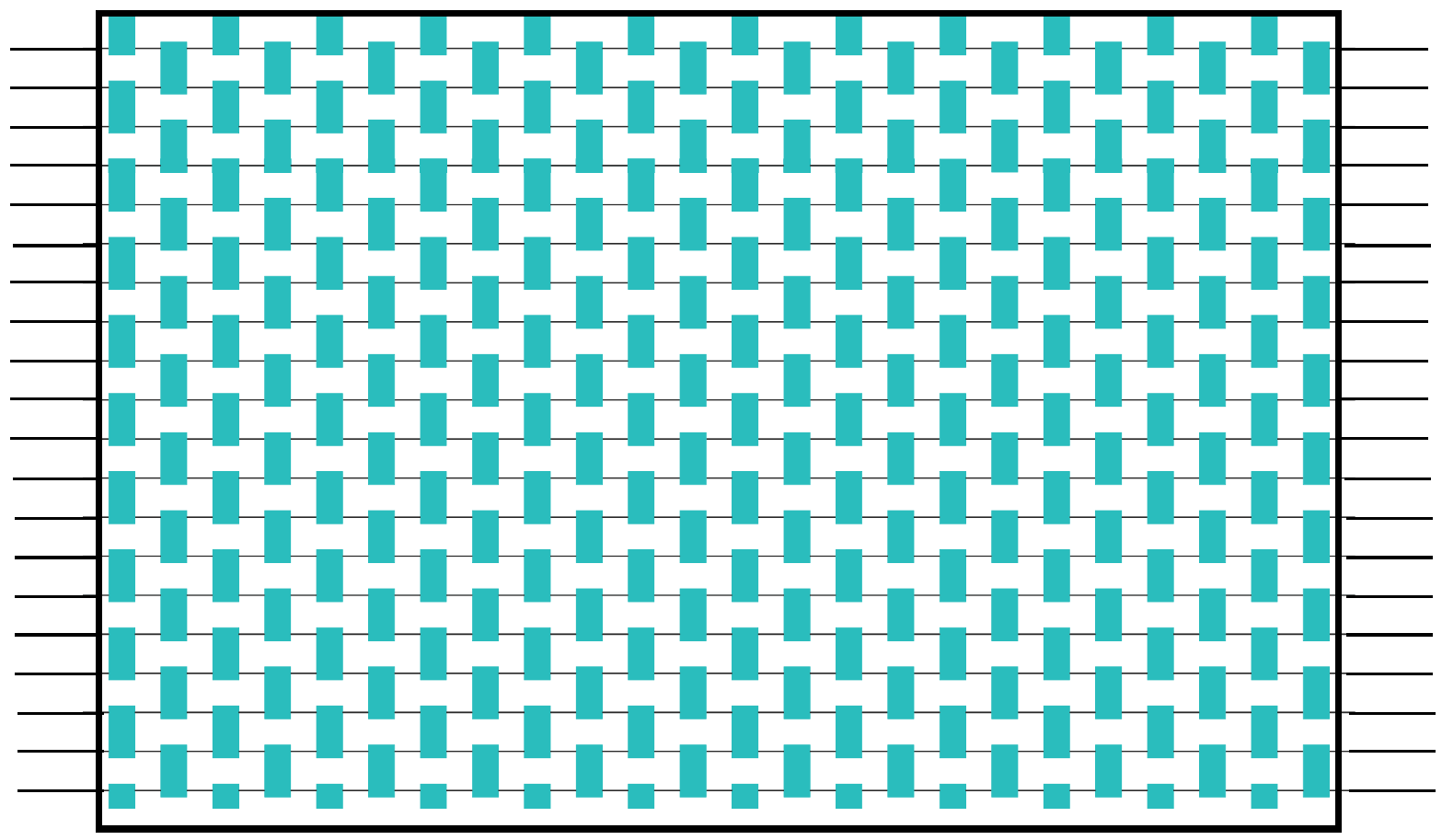}
\caption{\label{fig:flowchart} {\bf OPT finer and coarser descriptions.} The OPT is a {\em black-box approach}. It can be made more or less detailed, as in in the box on the left, and even at so fine a level that it is equivalent to a field-theoretical description, as in the right figure.}\label{f:bbox}
\end{center}
\end{figure}

Finally, the OPT is a {\em black-box approach}, where each test is described by a mathematical object which can be "actually achieved" by a very specific physical device. However, nobody forbids to provide a more detailed OPT realisation of the test, e.g. as in Fig. \ref{f:bbox}. 

\begin{figure}[h]
\begin{center}
\begin{tabular}{cc}
\includegraphics[width=5cm]{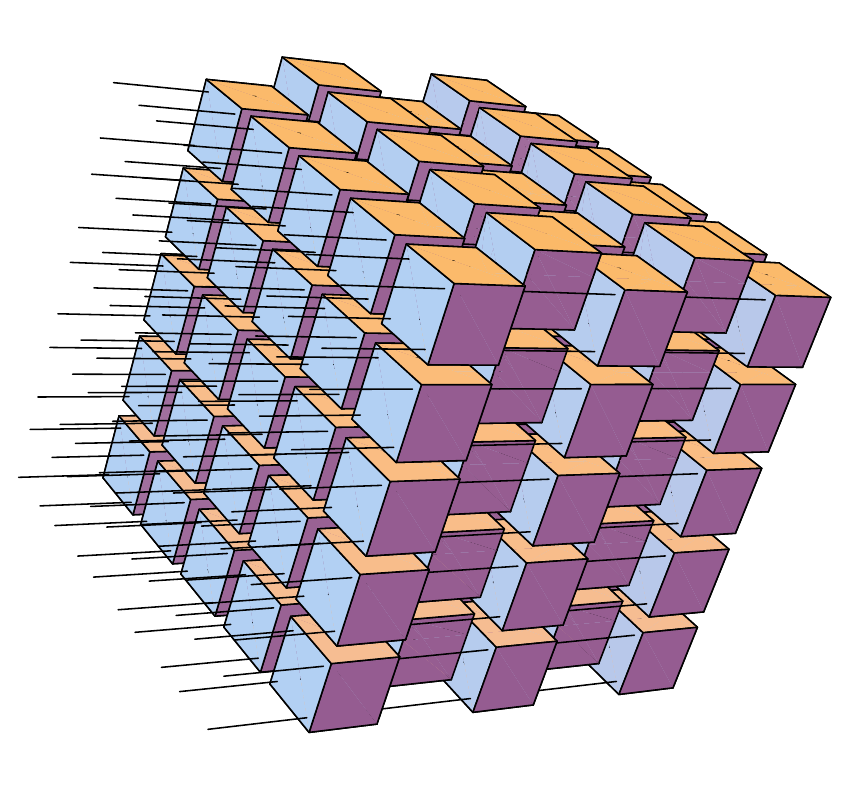}&
\end{tabular}
\caption{An (open) DAG whose topology is suitably representable in 3D.}\label{f:3DDAG}
\end{center}
\end{figure}

Notice that although any graph can be represented in 2D (e.g. using crossing of wires), one can more suitably design it in 3D (2D + in-out), as in Fig. \ref{f:3DDAG}.

 \subsubsection*{The OPT and the goal of science}
One can soon realise that the OPT framework precisely allows to express the most general goal of science, namely to connect objective facts happening (the events), devising a theory of such "connections" (the systems), thus allowing making predictions for future occurrences in terms of joint probabilities of events depending on their connections. 

One of the main methodologically relevant features of the OPT is that it makes perfectly distinct what is the "objective datum" from  what is "a theoretical element". What is {\em objective} is which tests are performed, and what is the outcome of each test. What is {\em theoretical} is the graph of connections between the tests, along with the mathematical representations of both systems and tests. The OPT framework dictates the rules that the mathematical description should satisfy, and the specific OPT gives the particular mathematical representation of systems and tests/events and of their compositions (in sequence and in parallel) to build up the CDAG. 

\subsubsection*{The OPT and the scientific method}
One of the main rules of the scientific method is to have a clearcut  distinction between what is experimental and what is theoretical. Though this would seem a trivial statement, such a confusion happens to be often the source of disagreement between scientists. Though the description of the apparatus is generally intermingled with theoretical notions, the pure experimental datum must have a conventionally defined "objectivity status", corresponding to "openly known" information, namely shareable by any number of different observers. Then both the theoretical language and the framework must reflect the theory-experiment distinction, by indicating explicitly which notions are assigned the objectivity status. Logic, with the Boolean calculus of events, is an essential part of the language, and Probability Theory can be regarded as an extension of logic, assigning probabilities to events. The notion that is promoted to the objectivity status is that of "outcome of a test", announcing which event of a given test has occurred. The OPT framework thus represents an extension of probability theory, providing a theoretical connectivity between events, the other theoretical ingredients being the mathematical descriptions of systems and tests.

 \subsubsection*{The OPT as a general information theory}

\begin{figure}[h]
\begin{center}
\begin{tabular}{cc}
\includegraphics[width=11.5cm]{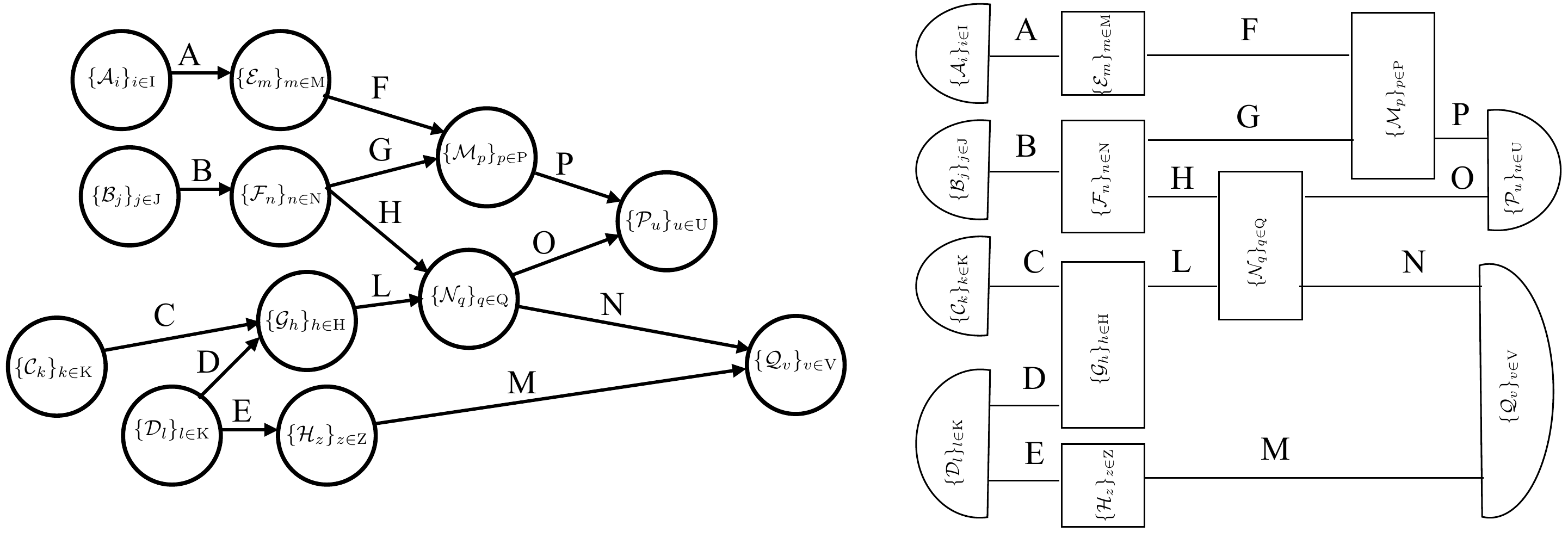}
\end{tabular}
\caption{\label{fig:flowchart} Equivalence between the CDAG and the quantum information circuit  or, equivalently, any run-diagram of a program.}\label{f:flow}
\end{center}
\end{figure}
We can immediately realise that a CDAG is exactly the same graph of a quantum circuit as it is drawn in quantum information science. The quantum circuit, in turn, can be interpreted as the run-diagram of a program, where each test represents a subroutine, and the wires represent the registers through which the subroutines communicate data. Indeed, the OPT can be regarded as the proper framework for information science in general.

For a recent complete presentation of the OPT framework, the reader is addressed to the the work     \cite{purification,CUPDCP} or the more recent thorough presentation      \cite{Chiribella2016}.
\vfill\clearpage\newpage
\section*{Notation and abbreviations}
\begin{table}[h]\renewcommand{\arraystretch}{.9}
\begin{tabular}{|l|l|}
\hline
$\Bnd^+(\sH)$  &bounded positive operators over $\sH$\\
$\CP_{\leq}$ &trace-non increasing completely positive map\\
$\CP_{=}$ &trace-preserving completely positive map\\
$\sH$ & Hilbert space over $\Cmplx$\\
$\Cone(\Set)$ &conic hull of $\Set$\\
$\Cone_{\leq 1}(\Set)$ &convex hull of $\{\Set\cup{0}\}$\\
$\Conv(\Set)$ &convex hull of $\Set$\\
$\Eff(\rA)$& set of effects of system $\rA$ \\
$\Eff_1(\rA)$& set of deterministic effects of system $\rA$ \\
$\Mrkv_{\leq}$ &normalization-non-incressing right-stochastic Markov matrices\\
$\Mrkv_1$ &normalization-preserving right-stochastic Markov matrices\\
$\Prm(n)$ &$n\times n$ permutation matrices\\
$(\Reals^n)^+_{\leq1}$&  $\{\vx\in\Reals^{n_\rA}|\vx\geq 0, \vx\leq\bf1\}$: simplex ${\bf S}^{n_\rA+1}$ \\
$(\Reals^n)^+_{1}$& $\{\vx\in\Reals^{n_\rA}|\vx\geq 0, \n{\vx}_1=1\}$: simplex ${\bf S}^{n_\rA}$ \\
$\St(\rA)$ &set of states of system $\rA$\\ 
$\St_1(\rA)$& set of deterministic states of system $\rA$\\ 
$\T(\sH)$ &trace-class operators over $\sH$\\
$\T^+(\sH)$  &trace-class positive operators over $\sH$\\
$\T_{\leq1}^+(\sH)$ &positive sub-unit-trace operators over $\sH$\\
$\T_{=1}^+(\sH)$ &positive unit-trace operators over $\sH$\\
$\Trn(\rA\to\rB)$& set of transformations from system $\rA$ to system $\rB$  \\
$\Trn_1(\rA\to\rB)$& set of deterministic transformations from system $\rA$ to system $\rB$\\
$\Uset(\sH)$ &unitary group over $\sH$\\
\hline
& {\bf Special cases corollaries}\\
\hline
&$\T(\Cmplx)=\Cmplx$,\;\;   $\T^+(\Cmplx)=\Reals^+$,\;\; $\T_{\leq1}^+(\Cmplx)=[0,1]$,\;\; $\T_{=1}^+(\Cmplx)=\{1\}$\\
&$\CP(\T(\sH)\to\T(\Cmplx))=\P(\T(\sH)\to\T(\Cmplx))=\{\Tr[\cdot E],\, E\in\Bnd^+(\sH)\}$
\\
&$\CP(\T(\Cmplx)\to\T(\sH))=\P(\T(\Cmplx)\to\T(\sH))=\T^+(\sH)$
\\
&$\CP_{\leq}(\T(\Cmplx)\to\T(\sH))\equiv\T^+_{\leq 1}(\sH)$\\
&$\CP_{\leq}(\T(\sH)\to\T(\Cmplx))\equiv\{\epsilon(\cdot)=\Tr[\cdot E],\, 0\leq E\leq I\}$\\
&$\Mrkv_{\leq}(n,1)=(\Reals^n)^+_{\leq1}$\\
&$\Mrkv _{1}(n,1)=(\Reals^n)^+_{=1}$\\
\hline
& {\bf Abbreviations}\\\hline
CT& Classical Theory\\
OPT& Operational Probabilistic Theory\\
PR boxes & Popescu-Rohrlich boxes\\
QT& Quantum Theory\\
\hline
\end{tabular}
\caption{Notation, special-cases corollaries, and common abbreviations.}\label{tnotat}
\end{table}

\newpage
\section*{Quantum theory}

A minimal mathematical axiomatisation of Quantum Theory as an OPT is provided in Table \ref{tabminimalQ}. For an OPT we need to provide the mathematical description of  systems, their composition, and transformations from one system to another. Then all rules of compositions of transformations and their respective systems are provided by the OPT framework. The reader who is not familiar with such framework can simply use the intuitive construction of quantum circuits. In Table \ref{tabmainthmsQ} we report the main theorems following from the axioms. The reader interested in the motivations for the present  axiomatization is addressed to Ref. \cite{DAriano2020}.
 \begin{table}[h]\renewcommand{\arraystretch}{.9}
\begin{center}
\begin{tabular}{|r| c|l|}
\hline
\multicolumn{3}{c}{\textbf{Quantum Theory}} \\
\hline
system &$\rA$& $\sH_\rA$\\
\hline
system composition &$\rA\rB$& $\sH_{\rA\rB}=\sH_\rA\otimes\sH_\rB$\\
\hline
transformation & $\tT\in\Trn(\rA\to\rB)$ & $\tT\in\CP_{\leq}(\T(\sH_\rA)\to\T(\sH_\rB))$
\\
\hline
Born rule &$p(\tT)=\Tr\tT$& $\tT\in\Trn(\rI\to\rA)$\\
\hline
\end{tabular}
\smallskip
\caption{{\bf Mathematical axiomatisation of Quantum Theory.} As given in the table, in Quantum Theory to each system $\rA$ we associate a Hilbert space over the complex field $\sH_\rA$. To the composition of systems $\rA$ and $\rB$ we associate the tensor product of Hilbert spaces $\sH_{\rA\rB}=\sH_{\rA}\otimes\sH_{\rB}$.  Transformations from system $\rA$ to $\rB$ are described by trace-nonincreasing completely positive (CP) maps from traceclass operators on $\sH_\rA$ to traceclass operators on $\sH_\rB$. Special cases of transformations are those with input trivial system $\rI$ corresponding  to states, whose trace is the preparation probability, the latter providing an efficient Born rule from which one can derive all joint probabilities of any combination of transformations. Everything else is simply special-case corollaries and one realisation theorem: these are reported in Table \ref{tabmainthmsQ}. 
}\label{tabminimalQ}
\end{center}
\end{table}
\begin{table}[h]\renewcommand{\arraystretch}{.9}
\begin{center}
\begin{tabular}{|r| c|l|}
\hline
\multicolumn{3}{c}{\textbf{Quantum theorems}} \\
\hline
trivial system &$\rI$& $\sH_\rI=\Cmplx$\\
\hline
reversible transf. & $\tU=U\cdot U^\dag$ & $U\in\Uset(\sH_\rA)$\\
\hline
determ. transformation & $\tT\in\Trn_1(\rA\to\rB)$ & $\tT\in\CP_{\leq1}(\T(\sH_\rA)\to\T(\sH_\rB))$\\
\hline
parallel composition &  $\tT_1\in\Trn(\rA\to\rB)$, $\tT_2\in\Trn(\rC\to\rD)$ &$\tT_1\otimes\tT_2$ \\
\hline
sequential composition &  $\tT_1\in\Trn(\rA\to\rB)$, $\tT_2\in\Trn(\rB\to\rC)$ &$\tT_2\tT_1$ \\
\hline
\multirow{4}{*}{states}& $\rho\in\St(\rA)\equiv\Trn(\rI\to\rA)$ & $\rho\in\T^+_{\leq1} (\sH_\rA)$\\
\cline{2-3}
  & $\rho\in\St_1(\rA)\equiv\Trn_1(\rI\to\rA)$ & $\rho\in\T^+_{=1} (\sH_\rA)$\\
\cline{2-3}
          & $\rho\in\St(\rI)\equiv\Trn(\rI\to\rI)$ & $\rho\in[0,1]$\\
\cline{2-3}
          & $\rho\in\St_1(\rI)\equiv\Trn(\rI\to\rI)$ & $\rho=1$\\
\hline
\multirow{2}{*}{effects}& $\epsilon\in\Eff(\rA)\equiv\Trn(\rA\to\rI)$ & $\epsilon(\cdot)=\Tr_\rA[\cdot E],\; 0\leq E\leq I_A$\\
\cline{2-3}
  & $\epsilon\in\Eff_1(\rA)\equiv\Trn_1(\rA\to\rI)$ & $\epsilon=\Tr_\rA$\\
\hline
\shortstack{\\Transformations as \\ unitary interaction\\ +\\ von Neumann-Luders}
    &
 $ \!\!\!\!\!\begin{aligned}
      \Qcircuit @C=1em @R=.7em @! R {
        &\poloFantasmaCn{\rA}\qw&\gate{\tT_i}&\poloFantasmaCn{\rB}\qw&\qw}
    \end{aligned} =\!\!\!\!\!\!\!\!\!\!\!\!
    \begin{aligned}
  \Qcircuit
    @C=1em @R=.7em @! R {
&\poloFantasmaCn{\rA}\qw&\multigate{1}{\tU}  &\poloFantasmaCn{\rB}\qw&\qw\\
\prepareC{\sigma}  &\poloFantasmaCn{\rF}\qw&\pureghost{\tU}\qw&\poloFantasmaCn{\rE}\qw&\measureD{\tP_i}
}
\end{aligned}$\!\!
& 
  \shortstack{$\tT_i\rho=\Tr_\rE[U(\rho\otimes\sigma)U^\dag(I_\rB\otimes P_i)]$}
\\
\hline
\end{tabular}
\smallskip
\caption{{\bf Corollaries and a theorem of Quantum Theory, starting from Table \ref{tabminimalQ} axiomatization.}
The first corollary states that the trivial system $\rI$ in order to satisfy the composition rule $\rI\rA=\rA\rI=\rA$ must be associated to the one-dimensional Hilbert space $\sH_\rI=\Cmplx$, since it is the only Hilbert space which trivializes the Hilbert space tensor product. The second corollary states that the reversible transformations are the unitary ones. The third corollary states that the deterministic transformations are the trace-preserving ones. Then the fourth and fifth corollaries give the composition of transformations in terms of compositions of maps. We then have four corollaries about states: 1) states are transformations starting from the trivial system and, as such, are positive operators on the system Hilbert space, having trace bounded by one; 2) the deterministic states correspond to unit-trace positive operator;  3) the states of trivial system are just probabilities; 4) The only trivial system deterministic state is the number 1. We then have two corollaries for effects, as special cases of transformation toward the trivial system: 1) the effect is represented by the partial trace over the system Hilbert space of the multiplication with a positive operator bounded by the identity over the system Hilbert space; 2) the only deterministic effect is the partial trace over the system Hilbert space. Finally, we have the realization theorem for transformations in terms of unitary interaction $\tU=U\cdot U^\dag$ with an environment $\rF$ and a projective effect-test $\{\tP_i\}$ over environment $\rE$, with $\tP_i=P_i\cdot P_i$, $\{P_i\}$ being a complete set of orthogonal projectors.
}\label{tabmainthmsQ}
\end{center}
\end{table}
\vfill

\vfill\clearpage\newpage
\section*{Classical theory}

\begin{table}[h]\renewcommand{\arraystretch}{.9}
\begin{center}
\begin{tabular}{|r| c|l|}
\hline
\multicolumn{3}{c}{\textbf{Classical Theory}} \\
\hline
system &$\rA$& $\Reals^{n_\rA}$\\
\hline
system composition &$\rA\rB$& $\Reals^{n_{\rA\rB}}=\Reals^{n_\rA}\otimes \Reals^{n_\rB}$\\
\hline
transformation & $\tT\in\Trn(\rA\to\rB)$ & $\tT\in\Mrkv_{\leq}(\Reals^{n_\rA}, \Reals^{n_\rB})$\\
\hline
\end{tabular}\smallskip
\caption{{\bf Mathematical axiomatisation of Classical Theory.} To each system $\rA$ we associate a real Euclidean space $\Reals^{n_\rA}$. To composition of systems $\rA$ and $\rB$ we associate the tensor product spaces $\Reals^{n_\rA}\otimes \Reals^{n_\rB}$.  Transformations from system $\rA$ to system $\rB$ are described by substochastic Markov matrices from the input  space to the output  space. Everything else are simple special-case corollaries: these are reported in Table \ref{tabmainthmsC}. 
}\label{tabminimalC}
\end{center}
\end{table}
\begin{table}[h]\renewcommand{\arraystretch}{.9}
\begin{center}
\begin{tabular}{|r| c|l|}
\hline
\multicolumn{3}{c}{\textbf{Classical theorems}} \\
\hline
trivial system &$\rI$& $\sH_\rI=\Reals$\\
\hline
reversible transformations & $\tP$  & $\tP\in\Prm(n_\rA)$\\
\hline
transformation & $\tT\in\Trn_\leqslant (\rA\to\rB)$ & $
\tT\in\Mrkv_\leqslant(\Reals^{n_\rA},\Reals^{n_\rB})$\\
\hline
determ. transformation & $\tT\in\Trn_1(\rA\to\rB)$ & $
\tT\in\Mrkv_1(\Reals^{n_\rA},\Reals^{n_\rB})$\\
\hline
parallel composition &  $\tT_1\in\Trn(\rA\to\rB)$, $\tT_2\in\Trn(\rC\to\rD)$ &$\tT_1\otimes\tT_2$ \\
\hline
sequential composition &  $\tT_1\in\Trn(\rA\to\rB)$, $\tT_2\in\Trn(\rB\to\rC)$ &$\tT_2\tT_1$ \\
\hline
\multirow{4}{*}{states}& $\vx\in\St(\rA)\equiv\Trn(\rI\to\rA)$ & $\vx\in(\Reals^{n_\rA})^+_{\leq1}$\\
\cline{2-3}
& $\vx\in\St_1(\rA)\equiv\Trn_1(\rI\to\rA)$ & $\vx\in(\Reals^{n_\rA})^+_{=1}$\\
\cline{2-3}
          & $p\in\St(\rI)\equiv\Trn(\rI\to\rI)$ & $p\in[0,1]$\\
\cline{2-3}
          & $p\in\St_1(\rI)\equiv\Trn(\rI\to\rI)$ & $p=1$\\
\hline
\multirow{2}{*}{effects}& $\epsilon\in\Eff(\rA)\equiv\Trn(\rA\to\rI)$ & $\epsilon(\cdot)=\cdot \vx,\; {\bf 0}\leq \vx\leq {\bf 1}$\\
\cline{2-3}
  & $\epsilon\in\Eff_1(\rA)\equiv\Trn_1(\rA\to\rI)$ & $\epsilon=\cdot {\bf 1}$\\
\hline
\end{tabular}
\caption{{\bf Main theorems of Classical Theory, starting from axioms in Table \ref{tabminimalC}}.
The first corollary states that the trivial system $\rI$ in order to satisfy the composition rule $\rI\rA=\rA\rI=\rA$ must be associated to the one-dimensional space $\Reals$, since it is the only real linear space that trivialises the tensor product. The second corollary states that the reversible transformations are the permutation matrices. The third states that transformations are substochastic Markov matrices. The fourth states that the deterministic transformations are stochastic Markov matrices. Then the fifth and sixth corollaries give the composition of transformations in terms of composition of matrices. We then have four corollaries about states: 1) states are transformations starting from the trivial system and, as such, are sub-normalized probability vectors (vectors in the positive octant with sum of elements bounded by one; 2) the deterministic states correspond to normalised probability vectors;  3) the case of trivial output-system correspond to just probabilities; 4) The only trivial output-system deterministic state is the number 1. We then have two corollaries for effects, as special cases of transformation toward the trivial system: 1) the effect is represented by scalar product with a vector with components in the unit interval; 2) the only deterministic effect is the scalar product with the vector with all unit components.
}\label{tabmainthmsC}
\end{center}
\end{table}


\vfill\newpage
\begin{acknowledgement}
The authors acknowledge helpful conversations and encouragement by Don Hoffman, and interesting discussions with Chris Fields. Giacomo Mauro D'Ariano acknowledges interesting discussions with Ramon Guevarra Erra. 

This work has been sponsored by Elvia and Federico Faggin foundation through Silicon Valley Community Foundation, Grant 2020-214365 {\em The observer: an operational theoretical approach.} For oral sources see also the Oxford podcast http://podcasts.ox.ac.uk/mauro-dariano-awareness-operational-theoretical-approach.
\end{acknowledgement}


\bibliographystyle{unsrt} 
\bibliography{Qconscious-Springer-2.bbl}
\end{document}